\documentclass[letterpaper,twocolumn,10pt]{article}
\usepackage{preprint}
\usepackage[margin=0.5in,bottom=0.75in,columnsep=0.25in]{geometry}
\usepackage[dvipsnames,table,xcdraw]{xcolor}
\usepackage[bookmarks, colorlinks, breaklinks]{hyperref}
\hypersetup{linkcolor=red, citecolor=cyan}

\usepackage{authblk}

\usepackage[T1]{fontenc}
\usepackage{helvet}

\usepackage{amsmath}
\usepackage{amssymb}
\usepackage{mathtools}
\usepackage{graphicx}
\usepackage[
  separate-uncertainty=true,
  multi-part-units=single,
  list-units=repeat,
  detect-weight,
  mode=text
]{siunitx}

\usepackage[tiny,compact]{titlesec}
\titleformat{\abstract}[runin]{\bfseries}{\theabstract}{1em}{}
\titleformat{\section}{\bfseries}{\thesection}{1em}{}
\titleformat{\subsection}[runin]{\bfseries}{\thesubsection}{1em}{}
\titleformat{\subsubsection}[runin]{\bfseries}{\thesubsubsection}{1em}{}
\titleformat{\paragraph}[runin]{\itshape}{\thesubsection}{1em}{}

\usepackage[font={small}]{caption}
\usepackage[capitalize]{cleveref}
\crefname{section}{Sec.}{Secs.}
\Crefname{section}{Section}{Sections}
\Crefname{table}{Table}{Tables}
\crefname{table}{Tab.}{Tabs.}

\usepackage[numbers,compress]{natbib}
\usepackage{optidef}
\usepackage{nicefrac}
\usepackage{booktabs}

\usepackage{xspace}

\providecommand{\sct}[1]{{\sc \texttt{#1}}}

\newcommand{\Dcorr}{\sct{Dcorr}\xspace}
\newcommand{\ttest}{\sct{t-test}\xspace}
\newcommand{\hotellings}{\sct{Hotelling}\xspace}
\newcommand{\mw}{\sct{MW}\xspace}
\newcommand{\anova}{\sct{ANOVA}\xspace}
\newcommand{\manova}{\sct{MANOVA}\xspace}
\newcommand{\ie}{\sct{IE}\xspace}
\newcommand{\omni}{\sct{Omni}\xspace}

\newcommand{\rdpg}{\sct{RDPG}\xspace}

\newcommand{\sbm}{\sct{SBM}\xspace}
\newcommand{\tnorm}{\mathcal{TN}\xspace}
\newcommand{\bern}{\mathrm{Bernoulli}\xspace}
\newcommand{\cmds}{\sct{cMDS}\xspace}

\newcommand{\A}{\mathbf{A}\xspace}
\newcommand{\Aij}{\A_{ij}\xspace}
\newcommand{\B}{\mathbf{B}\xspace}
\renewcommand{\P}{\mathbf{P}\xspace}
\newcommand{\Pij}{\P_{ij}\xspace}
\newcommand{\X}{\mathbf{X}\xspace}
\newcommand{\R}{\mathbb{R}\xspace}
\newcommand*{\T}{{^\intercal}}

\newcommand{\pperp}{\perp \!\!\!\perp}

\usepackage{newfloat}
\DeclareFloatingEnvironment[name={Supplementary Figure}, fileext=lsf, listname={List of Supplementary Figures}]{suppfigure}

\title{Multiscale Comparative Connectomics}
\author[1]{Vivek Gopalakrishnan$^\dagger$}
\author[1]{Jaewon Chung}
\author[2]{Eric Bridgeford}
\author[1]{Benjamin D. Pedigo}
\author[3]{Jes\'us Arroyo}
\author[4]{Lucy Upchurch}
\author[4,5]{G. Allan Johnson}
\author[6]{Nian Wang}
\author[7]{Youngser Park}
\author[7,8]{Carey E. Priebe}
\author[1,2,7,8]{Joshua T. Vogelstein}
\affil[1]{Department of Biomedical Engineering, Johns Hopkins University}
\affil[2]{Department of Biostatistics, Johns Hopkins University}
\affil[3]{Department of Statistics, Texas A\&M University}
\affil[4]{Center for In Vivo Microscopy, Department of Radiology, Duke University}
\affil[5]{Department of Biomedical Engineering, Duke University}
\affil[6]{Department of Radiology and Imaging Sciences, Indiana University School of Medicine}
\affil[7]{Center for Imaging Science, Johns Hopkins University}
\affil[8]{Department of Applied Mathematics and Statistics, Johns Hopkins University}
\affil[$\dagger$]{\it Correspondence to: \href{mailto:vivekg@mit.edu}{vivekg@mit.edu}}

\begin{document}

\twocolumn[{
\renewcommand\twocolumn[1][]{#1}
\maketitle
\begin{abstract}
The connectome, a map of the structural and/or functional connections in the brain, provides a complex representation of the neurobiological phenotypes on which it supervenes. This information-rich data modality has the potential to transform our understanding of the relationship between patterns in brain connectivity and neurological processes, disorders, and diseases. However, existing computational techniques used to analyze connectomes are oftentimes insufficient for interrogating multi-subject connectomics datasets: many current methods are either solely designed to analyze single connectomes or leverage heuristic graph statistics that are unable to capture the complete topology of multiscale connections between brain regions. To enable more rigorous connectomics analysis, we introduce a set of robust and interpretable effect size measures motivated by recent theoretical advances in random graph models. These measures facilitate simultaneous analysis of multiple connectomes across different scales of network topology, enabling the robust and reproducible discovery of hierarchical brain structures that vary in relation to phenotypic profiles. In addition to explaining the theoretical foundations and guarantees of our algorithms, we demonstrate their superiority over current state-of-the-art connectomics methods through extensive simulation studies and real-data experiments. Using a set of high-resolution connectomes obtained from genetically distinct mouse strains (including the BTBR mouse---a standard model of autism---and three behavioral wild-types), we illustrate how our methods successfully uncover latent information in multi-subject connectomics data and yield valuable insights into the connective correlates of neurological phenotypes that other methods do not capture. The data and code necessary to reproduce the analyses, simulations, and figures presented in this work are available at \url{https://github.com/neurodata/MCC}.
\end{abstract}
\vspace{1em}
}]

\section{Introduction}

Understanding how patterns in brain connectivity give rise to observable biological phenotypes is a central pursuit in neuroscience. Derived from neuroimaging data, the connectome (a graphical representation of neural connections) has recently become an invaluable modality for such analyses, allowing researchers to represent organisms' brains as networks and understand nervous system organization with graph theoretical methods \citep{bullmoreComplexBrainNetworks2009}. Successfully associating biological phenotypes with organizational variation in the connectome will enable the identification of neurological structures and circuits that drive cognitive function \citep{bullmoreBrainGraphsGraphical2011a}. However, to fully realize the promise of the connectome, new statistical graph theory methods that are principled, robust, and reproducible are required for the analysis of this nascent and highly-complex datatype \citep{craddockImagingHumanConnectomes2013a}.

From a mathematical perspective, a connectome can be modeled as a network (also referred to as a graph) of the interactions between brain regions \citep{spornsHumanConnectomeStructural2005}. In this network, vertices represent disjoint regions of the brain, and edges represent the connections between these regions \citep{vogelsteinConnectalCodingDiscovering2019a}. From a neuroscientific perspective, the connectome can be further described at multiple hierarchical scales of network topology \citep{kaiserTutorialConnectomeAnalysis2011}: at the extremes are the \textit{local scale}, characterized by the features of individual edges and vertices in the connectome, and the \textit{global scale}, characterized by global patterns in brain connectivity and often quantified by a variety of graph statistics \citep{rubinovComplexNetworkMeasures2010}; the intermediate \textit{regional scale} focuses on the interactions between distinct subsets of brain regions (known as communities or blocks), comprising subgraphs of the connectome \citep{wangSignalSubgraphEstimation2018, vogelsteinGraphClassificationUsing2013}. Simultaneously considering the local, regional, and global scales of network topology provides a multiscale view of patterns in neural connectivity.

The fundamental goal of multi-subject connectomics is to identify the multiscale patterns in brain network architecture that differ across phenotypic groups \citep{vandenheuvelComparativeConnectomics2016a}. However, given the inherent structure of the connectome, special care needs to be taken when designing analytical methods. The naive application of classical statistical tests to network-valued data can sometimes produce misleading results. For example, many methods use graph statistics (e.g., clustering coefficient, degree centrality, etc.) to characterize differences in the connectivity patterns within connectomes; however, recent studies have shown that no set of graph statistics can comprehensively describe network topology, as networks with wildly different structures can produce identical graph statistics \citep{chenSameStatsDifferent2019, chungStatisticalConnectomics2020a}. That is, the one number summaries of edge-, vertex-, and community-scale connectivity provided by graph statistics often fail to comprehensively describe neurological topology. This is not to say the graph statistics are not a useful analytic tool (they are often intuitive and logical descriptors of individual connectomes), but rather that they are often insufficient when building statistical tests for comparative connectomics. In multiple simulation studies and real-data experiments, we demonstrate that tests based on graph statistics do not always perform accurate inference, failing to recapitulate known neurological phenomena from connectomes.

\begin{figure*}[t]
\centering
\includegraphics[width=\linewidth]{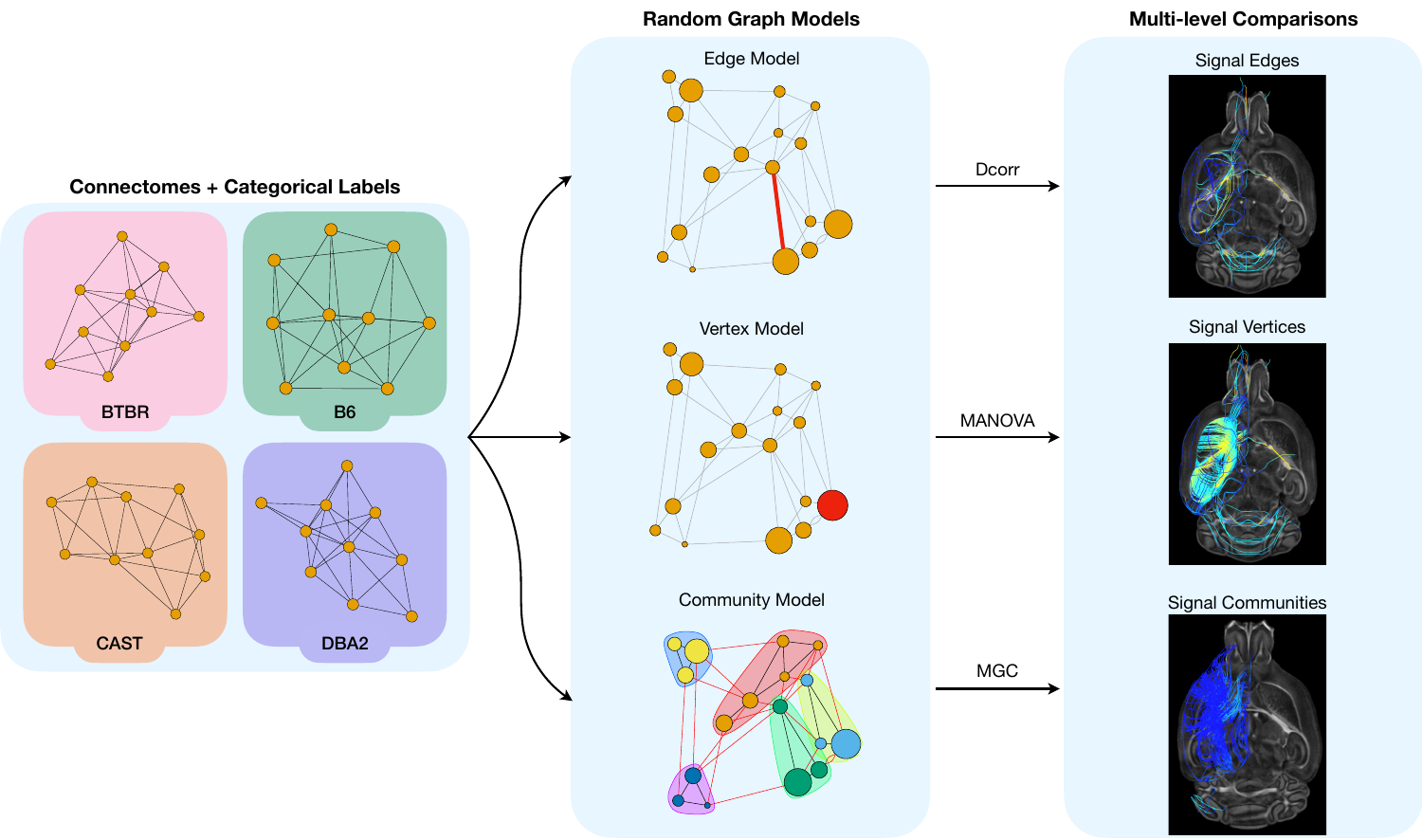}
\caption{Overview of the statistical framework for multi-level inference in multi-subject connectomics. (\textit{Left}) All connectomes are defined on a common set of vertices and have an associated categorical label. Illustrative networks from four mouse lines whose connectomes we analyze in this work are shown. (\textit{Center}) Connectomes are fit to random graph models that are specifically chosen to model the variation in a network at a given topological level. (\textit{Right}) Using the categorical labels, $k$-sample effect size measures are applied to the estimated parameters from each model, yielding a set of the signal edges, vertices, and communities across the groups in a given multi-subject connectomics dataset. Finally, we visualize the strongest signal component at each scale using tractography.}
\label{fig:pipeline}
\end{figure*}

To overcome these limitations and enable the rigorous interrogation of multi-subject connectomics datasets, we present a set of network-based effect size measures that provide insight across topological scales of the connectome (\Cref{fig:pipeline}). Predicated on recent advances in the theory of random graph models, our methods can be used to identify neurobiological structures within connectomes that are connectively different across multiple categorical or dimensional phenotypes, which we term \textit{signal} components. Specifically, these effect size measures can be used to discover signal edges, vertices, and communities within populations of connectomes defined on the same vertex set and are appropriate for analyzing connectomes estimated from either structural or functional neuroimaging data. We formulate these methods as $k$-sample effect size measures, enabling comparisons of connectomes from more than two distinct phenotypic groups. Finally, we show how these methods can be aggregated across scales, helping to overcome the limitations of multiple hypothesis testing that is inherent to connectomics data.

Modeling connectomes with random graph models allows us to mathematically characterize patterns in network structure, as well as to account for noise within and across samples. Such models have many interpretable and provable properties, which we leverage to formulate principled effect size measures for multi-subject connectomics data. We demonstrate the efficacy and utility of this connectomics paradigm by applying our multiscale measures to an open access dataset of ultrahigh-resolution structural mouse connectomes (derived at a spatial resolution 20,000 times greater than typical human connectomes) \citep{wangVariabilityHeritabilityMouse2020a}. Additionally, we compare the performance of our proposed tests to prevailing connectomics analysis strategies in extensive simulation studies, and show that the neurobiological insights of our measures of connectivity are orthogonal to existing measures through information-theoretic comparisons. In total, our proposed methods enable principled interrogation of the local, regional, and global scales of network topology, providing researchers with novel tools for testing neurobiological hypotheses in multisubject connectomics datasets.

\section{Preliminaries}

\subsection{Graph theory notation.}
Networks (or graphs) are convenient mathematical objects for representing connectomes. A network $\mathcal G$ consists of the ordered pair $(V, E)$, where $V$ is the set of vertices and $E$ is the set of edges.
The set of vertices can be represented as $V = \{ 1, 2, \dots, n \}$, where $|V| = n$ is the number of vertices.
The set of edges is a subset of all possible connections between vertices (i.e., $E \subseteq V \times V$).
We say the tuple $(i, j) \in E$ if there exists an connection between vertex $i$ and vertex $j$. In many connectomics datasets, edges have associated edge weights, real-valued numbers that encode quantitative information about a connection between two vertices. The interpretation of the edge weight is dependent on the imaging modality used to measure the connectome. For example, the edge weights in structural connectomes are non-negative integers that represent the number of neuronal fibers that traverse from one region of the brain to another \citep{kiarHighThroughputPipelineIdentifies2017}. Every connectome has an associated weighted adjacency matrix $\mathbf A \in \R_{\geq 0}^{n \times n}$ where $\mathbf A_{ij}$ denotes the weight of the edge $(i, j) \in E$.

\subsection{Random graph models.}
\label{sec:models}
Statistical modeling of connectomics data enables the principled analysis of these high-dimensional, graph-valued data. Random graph models treat individual connectomes as random variables, enabling mathematical characterization of network structure and accounting for noise within and across observed samples. Below, we present the three random graph models we use to construct our multi-level tests: 1) the Independent Edge (\ie) model; 2) the Random Dot Product Graph (\rdpg); and 3) the Stochastic Block Model (\sbm). Each model is designed to characterize a particular topological level of the connectome (i.e., either its edges, vertices, or communities). Treating connectomes as random network-valued variables sampled from these random graph models enables the formulation of $k$-sample hypothesis tests that can be used to identify connective differences at multiple levels across numerous phenotypic profiles.

\paragraph{The edge model.}
In the Independent Edge (\ie) model, every possible edge $(i, j) \in V \times V$ is sampled from independent Bernoulli distributions, each parameterized by some edge-specific probability $p_{ij} \in [0, 1]$. These can be summarized using a matrix of edge-wise probabilities $\P \in [0, 1]^{n \times n}$, where $\Pij = p_{ij}$. In our formulation of the edge-wise hypothesis test (\S\ref{sec:formulations}), we consider the weighted \ie model to account for a network with weighted edges. For this variant, instead of treating each edge as a Bernoulli random variable, we sample each edge weight from a distribution $F_{ij}$ supported on the set of non-negative real numbers. In the weighted \ie model, $\P$ represents a matrix of univariate probability distributions modeling the weight of each edge in the connectome. We say a network is sampled from the model, i.e., $G \sim \text{\ie}(\P)$, if its adjacency matrix $\A$ has entries $\Aij \sim F_{ij}$ independently for every edge in the connectome. When estimating the $\P$ matrix for a weighted \ie model, we assume that all elements of $\P$ are from the same family of distributions.

\paragraph{The vertex model.}
The Random Dot Product Graph (\rdpg) is a member of the family of latent position random graphs, a class of models where the probability of a connection $p_{ij}$ is determined by the vertices, not the edges \cite{hoffLatentSpaceApproaches2002}. Under such models, each vertex $i \in V$ is associated with a \textit{latent position} $x_i$, which belongs to some \textit{latent space} $\mathcal{X}$. The probability of a connection between vertices $i$ and $j$ is given by a link function $\kappa: \mathcal{X} \times \mathcal{X} \mapsto [0, 1]$; 
that is, $p_{ij} = \kappa(x_i, x_j)$. In the \rdpg, the latent space $\mathcal{X}$ is a subspace of Euclidean space $\mathbb{R}^d$ and the link function is the dot product \cite{scheinermanModelingGraphsUsing2010}. Thus, in a $d$-dimensional \rdpg with $n$ vertices, the rows of the matrix $\X \in \mathbb{R}^{n \times d}$ encode the latent position of each vertex, and the matrix of edge-wise connection probabilities is given by $\P = \X\X^\T$. A network is sampled from the model $G \sim \text{\rdpg}(\X)$ if its adjacency matrix $\A$ has entries $\Aij \sim \bern(x_i \cdot x_j)$ independently for every edge $(i, j) \in V \times V$.

The modeling assumptions of the \rdpg make estimation of the latent positions analytically tractable. The estimation procedure we use is the omnibus embedding (\omni) \cite{athreyaStatisticalInferenceRandom2017a}, which jointly estimates the latent positions for every connectome in a dataset by mapping each vertex to a vector in $\mathbb R^d$ that corresponds to the vertex's latent position in a $d$-dimensional \rdpg.  A host of downstream machine learning tasks can be accomplished with this jointly embedded representation of a sample of connectomes, such as clustering or classification of vertices. Here, we use the embedding to formulate an effect size measure that can be used to identify vertices that are strongly associated with given phenotypes. Note that \omni is a member of a larger class of \textit{vertex embedding} algorithms, i.e., algorithms that map vertices to real-valued vectors. However, the vectors produced by other vertex embedding algorithms do not correspond to parameters in an \rdpg. We compare \omni to many other vertex embedding methods, including vertex-level graph statistics, in the Results. Our experiments demonstrate that the embeddings produced by \omni enable accurate statistical inference by identifying signal vertices with known neurobiological significance, whereas previously proposed vertex embedding methods do not.

\paragraph{The community model.}
In the Stochastic Block Model (\sbm), every vertex belongs exclusively to one of $K$ communities, which partition the vertex set \cite{hollandStochasticBlockmodelsFirst1983a}. The \sbm is a special case of the \rdpg in which all vertices from the same community have identical latent positions; that is, the connection probability is solely determined by community membership. A symmetric $K \times K$ community connectivity probability matrix $\B$ with entries in $[0,1]^{K \times K}$ governs the probability of an edge between two vertices given their community memberships. Community membership is determined by the vertex assignment vector $\vec{\tau} \in \{ 1, \dots, K \}^n$, which is either unknown and estimated from the data, or given \textit{a priori}. Because brain regions in different atlases are usually hierarchically grouped into superstructures and hemispheres, we assume that $\vec{\tau}$ is given. Thus, a network is sampled from the model, $G \sim \text{\sbm}(\B)$ with $\vec{\tau}$ given, if its adjacency matrix $\A$ has entries $\Aij \sim \bern(\B_{k_i k_j})$ where $\tau_i = k_i$ and $\tau_j = k_j$ for $(i, j) \in V \times V$, and $k_i, k_j \in \{1, \dots, K\}$. As with the \ie model, a weighted variant of the \sbm can be constructed by replacing the block probability $\B_{ij}$ with a probability distribution $F_{ij}$ for the entire block. 

\section{Methods}

\subsection{Effect size measures for multiscale inference.}
\label{sec:formulations}
We introduce our $k$-sample effect size measures using a formal statistical framework. Practical demonstrations of how these measures can be used to analyze real-world connectomics data are provided in the Results (\S\ref{sec:results}).

\paragraph{Identifying signal edges.}
The simplest approach for comparing connectomes is to treat them as a \textit{bag of edges} without considering interactions between the edges \cite{craddockImagingHumanConnectomes2013a}. Serially performing univariate statistical tests at each edge enables the discovery of \textit{signal edges} whose neurological connectivity differs across categorical or dimensional phenotypes. Using the \ie model, we assume that each connectome is sampled from a phenotype-conditional probability matrix; that is, we assume that for each phenotype in $\mathcal{Y} = \{ c_1, \dots, c_k \}$, there is an associated probability matrix in $\left\{ \mathbf{P}^{c_1}, \dots, \mathbf{P}^{c_k} \right\}$ from which all connectomes in that phenotype are sampled. For a given edge $(i, j)$, we assume the edge weight for each connectome has been independently and identically (i.i.d) sampled from the appropriate $\mathbf{P}$ matrix. Specifically, we assume that for every connectome in phenotype $c_y \in \mathcal Y$, the edge weight $\Aij \sim \mathbf{P}^{c_y}_{ij}$ i.i.d. Following this assumption, we formulate the following null and alternative hypotheses:
\begin{align*}
    H_0 & : \forall (y,y'):\quad \P^{c_y}_{ij} = \P^{c_{y'}}_{ij} \\
    H_1 & : \exists (y,y'):\quad \P^{c_y}_{ij} \neq \P^{c_{y'}}_{ij}
\end{align*}
That is, the null hypothesis is that weight distribution for a particular edge $(i, j)$ is identical across all phenotypic groups, whereas the alternative hypothesis that this distribution is different for at least one of the phenotypes.

To test the null hypothesis, any univariate statistical test can be employed since edge weights are themselves scalar random variables in the \ie model. For this task, since the entries of $\P$ are themselves distributions, we use Distance Correlation (\Dcorr), a previously established universally consistent nonparametric $k$-sample test for equality in distribution \cite{szekelyMeasuringTestingDependence2007, pandaHyppoComprehensiveMultivariate2020}. In the Simulations (\S\ref{sec:edgesim}), we demonstrate that \Dcorr is a more powerful test than commonly used non-parametric and Gaussian alternatives. 

As a final consideration, note that serial edge-wise testing requires corrections for an immense number of multiple comparisons. If the sample consists of directed connectomes, then the total number of tests is $n^2$; if the connectomes are undirected, then the total number of tests is $\binom{n}{2}$.

\paragraph{Identifying signal vertices.}
We test for differences in a given vertex's connectivity by comparing its latent position estimates from \omni across phenotypes. According to a Central Limit Theorem for \omni, these latent position estimates are universally consistent and asymptotically normal \cite{levinCentralLimitTheorem2017b}. This motivates our use of normal-theory inferential statistical tests to determine if the embedding of a given vertex is different across phenotypes. If the number of classes $k=2$, we use Hotelling's T-squared (\hotellings), a multivariate generalization of the \ttest; if $k>2$, we use one-way \manova with Pillai's trace as our test statistic. We formulate the following null and alternative hypotheses:
\begin{align*}
    H_0 & : \forall (y,y'):\quad \mu_i^{c_y} = \mu_i^{c_{y'}} \\
    H_1 & : \exists (y,y'):\quad \mu_i^{c_y} \neq \mu_i^{c_{y'}}
\end{align*}
where $\mu_i^{c_y}$ is the mean latent position of the vertex $i$ for the phenotype $c_y \in \mathcal{Y}$. This procedure results in a total of $n$ statistical tests, one for each vertex in the vertex set $V$.

\paragraph{Identifying signal communities.}
\label{sec:regional-detail}
Vertices in a connectome can be hierarchically organized into superstructures such as major brain regions and hemispheres. The interactions within and between these communities of vertices form connective circuits within the brain, and are more correlated with complex behavior and phenotypes than single edges or vertices. Therefore, interrogation of the \textit{regional level} and identification of signal communities is critical component of multi-level connectomics analysis.

Here, we use the \sbm to model the community structure of a connectome. We propose four approaches for describing the connectivity of a community, and an accompanying statistical procedure for each approach. Each subsequent approach provides an increasingly more holistic description of a community. For a given block $(i, j)$ in the \sbm, and for each phenotype, we posit the following tests:
\begin{enumerate}
    \item \textbf{Average Connectivity:} Compute the average number of nonzero edges (binarize using Otsu's method \cite{otsuThresholdSelectionMethod1979}), which is equivalent to the community connectivity probability $\B_{ij}$. We then use Pearson's chi-squared to determine if there is a significant difference in the connectivity probability across phenotypes.
    \item \textbf{Average Edge Weight:} Compute the average weight of all the edges, essentially treating the community as a single large edge. We then use \Dcorr to determine if there is a significant difference in the average edge weight across phenotypes.
    \item \textbf{Multivariate Binary:} Binarize the community using Otsu's method \cite{otsuThresholdSelectionMethod1979}, and vectorize the subgraph of the adjacency matrix corresponding to that community. We binarize to determine if the distribution of nonzero edges is different across phenotypes. Test for differences in this binary representation using \Dcorr.
    \item \textbf{Multivariate Weighted:} Vectorize the subgraph of the adjacency matrix corresponding to that community. Again, test for differences in this weighted representation using \Dcorr.
\end{enumerate}
If the connectomes in question have $K$ communities and are directed, this procedure results in $K^2$ comparisons. If the connectomes are undirected, this procedure results in $\binom{K}{2}$ comparisons.

\subsection{Correcting for multiple comparisons.}
\label{sec:correction}
Conventional atlases for multi-subject connectomics studies are typically composed of hundreds of vertices and tens of thousands of edges \cite{myersStandardizingHumanBrain2019}. Therefore, as demonstrated by the formulations in the Methods (\S\ref{sec:formulations}), the multiscale measures of effect size we introduce result in a large number of multiple comparisons. To limit the number of structures that are falsely identified as signal components, we consider methods for controlling the Family-Wise Error Rate (FWER), the probability of making at least one false discovery, and the False Discovery Rate (FDR), the proportion of discoveries expected to be false.

Well-established methods for controlling the FWER like the Bonferroni correction \cite{weissteinBonferroniCorrectionMathWorld2004} are amenable to applications in connectomics because they do not assume that the multiple hypotheses are independent of each other. Such independence would be a logically impossible condition for all but the simplest random graph models (e.g., weighted Erd\H{o}s--R\'enyi) as each edge is defined specifically by virtue of a dependence between pairs of vertices. However, the stringent definition of FWER makes the Bonferroni correction (and even its uniformly more powerful extension, the Holm--Bonferroni correction \cite{holmSimpleSequentiallyRejective1979}) overly conservative. By comparison, the FDR-controlling Benjamini--Hochberg correction is less conservative; however, to guarantee control of the FDR, the procedure assumes that the p-values from true null hypotheses (called \textit{null p-values}) are mutually independent of one another, as well as being independent of the non-null p-values \cite{benjaminiControllingFalseDiscovery1995}. When applied to connectomics data, which necessarily violates the independence assumption, Benjamini--Hochberg can result in overly liberal statistical corrections and higher rates of Type 1 Error \cite{efronSimultaneousInferenceWhen2008}. To err on the side of statistical caution, we use the Holm--Bonferroni correction as its lack of an independence assumption makes it valid for connectomics data. By adopting FWER control instead of FDR control, we greatly reduce the likelihood of our algorithms producing spurious results.

To ameliorate the conservative nature of FWER control, we leverage a unique advantage of our multiscale approach. Unlike single scale methods for comparative connectomics, our proposed effect size measures are able to \textit{borrow power} across scales of the connectome. For example, given the relatively small sample sizes in connectomics datasets, it is difficult to accurately estimate the importance of a particular edge (tens of subjects versus hundreds of thousands of edges). However, it is more feasible to estimate the importance of a particular vertex (tens of subjects versus hundreds of vertices). Then, if a given vertex is determined to be a signal vertex, some subset of the incident edges must also be signal edges. This allows inferences at coarser resolutions of network connectivity to inform inferences at finder resolutions, where the dimensionality is so high that traditional statistical hypothesis tests will not be sufficiently powerful to identify signal components.

\subsection{A multi-subject mouse connectome dataset for algorithmic validation.}
In the Results (\S\ref{sec:results}), we provide illustrative examples of how our multi-level hypothesis tests operate on real-world data by applying them to an open access dataset of whole-brain diffusion magnetic resonance imaging-derived connectomes from four mouse lines: BTBR T+ Itpr3tf/J (BTBR), C57BL/6J (B6), CAST/EiJ (CAST), and DBA/2J (DBA2) \cite{wangVariabilityHeritabilityMouse2020a}. The BTBR mouse strain is a well-studied model that exhibits core behavioral deficits that characterize autism spectrum disorders (ASD) in humans \cite{silvermanRepetitiveSelfGroomingBehavior2010, mcfarlaneAutismlikeBehavioralPhenotypes2008, scattoniUnusualRepertoireVocalizations2008}. Additionally, the BTBR mouse has significant neuroanatomical abnormalities including the complete absence of the corpus callosum, a band of nerve fibers connecting the left and right hemispheres of the brain \cite{ellegoodNeuroanatomicalAnalysisBTBR2013a, meyzaBTBRMouseModel2017}. {Therefore, we can use this dataset to determine if our tests (as well as other existing connectomics methods) will successfully recover this previously established neurobiological information.} The B6, CAST, and DBA2 mice are genetically distinct strains that do not exhibit ASD-like behaviors; they serve as wild-type behavioral controls in these experiments.

For each strain, connectomes were generated from eight age-matched mice ($N=8$ per strain with a sex distribution of four males and four females) using diffusion tensor imaging (DTI) tractography as described in the Methods (\S\ref{sec:tractography}). Improved imaging protocols and custom hardware were developed and implemented to solve existing technical issues with DTI, including an inability to resolve crossing and merging fibers (for more details, see \S2.2 of \cite{wangVariabilityHeritabilityMouse2020a}). Each connectome was parcellated using a symmetric Waxholm Space \cite{johnsonWaxholmSpaceImagebased2010, calabreseDiffusionMRITractography2015}, yielding a vertex set with a total of 326 regions of interest (ROIs) bilaterally distributed across the left and right hemispheres, and an undirected edge set with 52,975 edges. Within a given hemisphere, there were seven superstructures consisting of multiple ROIs, resulting in a total of 14 distinct communities in each connectome. Heatmaps of the average log-transformed adjacency matrix for each strain with hierarchical community and hemispheric labels are shown in Supplementary Figure \ref{fig:connectomes}.

\subsection{Tractography.}
\label{sec:tractography}
To visualize the neurological structures identified by our methods, deterministic tractograms were generated in DSI Studio using the generalized Q-sampling fiber tracking algorithm \cite{yeh_deterministic_2013}. For the specific parameter values used for tract generation, see \S2.6 of \cite{wangVariabilityHeritabilityMouse2020a}. A group average template was constructed from the 4 male mice per strain. A DTI diffusion scheme was used, and a total of 46 diffusion sampling directions were acquired. The b-value was 4000 \si{\second\per\milli\meter^2}. The in-plane resolution and the slice thickness were both 0.045 \si{\milli\meter}. The diffusion data were reconstructed in the MNI space using q-space diffeomorphic reconstruction \cite{yehNTU90HighAngular2011a} to obtain the spin distribution function \cite{fang-chengyehGeneralizedSamplingImaging2010}. A diffusion sampling length ratio of 1.25 was used. The restricted diffusion was quantified using restricted diffusion imaging \cite{yehMappingImmuneCell2017}.

\subsection{Data and code availability}
A real-world multi-subject mouse connectomics dataset analyzed in this paper was derived by Wang et al., and is described in further detail in the original publication \cite{wangVariabilityHeritabilityMouse2020a}. These data are freely available in \texttt{graspologic} (\url{https://github.com/microsoft/graspologic}) \cite{chungGraSPyGraphStatistics2019a}, an open-source Python package for statistical network analysis. All network-related analyses and simulations were performed using \texttt{graspologic}, and all multivariate hypothesis testing was performed using \texttt{hyppo} (\url{https://github.com/neurodata/hyppo}) \cite{pandaHyppoComprehensiveMultivariate2020}. Heatmaps were generated using \texttt{ComplexHeatmap} \cite{guComplexHeatmapsReveal2016a}. The code necessary to reproduce the analyses, simulations, and figures presented in this work are available in a series of Jupyter Notebooks at (\url{https://github.com/neurodata/MCC}).

\section{Results}
\label{sec:results}

\subsection{Interrogating the \textit{local scale}.}
\label{sec:micro}
Illustrative applications of our proposed local scale algorithms, along with comparisons to existing connectomics methods, are provided below. The mathematical formulations of these algorithms for identifying signal edges and signal vertices are provided in the Methods (\S\ref{sec:formulations}).

\subsubsection{Identifying signal edges.}
Univariate edge-wise testing provides an interpretable and computationally tractable method for identifying connective differences in specific edges across phenotypes. This is a well-established analytical approach in connectomics, and many classical statistical tests like one-way analysis of variance (\anova) and the Kruskal--Wallis \textit{H} test are commonly used for edge-wise testing \cite{dagenbachInsightsCognitionNetwork2019, craddockImagingHumanConnectomes2013a, varoquauxLearningComparingFunctional2013}. However, these are tests of equality in \textit{location}: that is, \anova (Kruskal--Wallis) tests whether the mean (median) weight of a particular edge is statistically different across phenotypes. If the distribution of an edge weight is different across phenotypes but the mean edge weight happens to be equal (for example, if one phenotype is bimodal for a particular weight), \anova and Kruskal--Wallis will fail to identify this signal edge. To account for this deficiency, we advocate for the use of Distance Correlation (\Dcorr)---a previously established universally consistent nonparametric test for equality in distribution \cite{szekelyMeasuringTestingDependence2007, pandaHyppoComprehensiveMultivariate2020}---to detect signal edges. \Dcorr measures the strength of both linear and nonlinear association between two random variables (in this case, the edge weight and the phenotypic label), and equals zero if and only if the random variables are independent \cite{szekelyMeasuringTestingDependence2007}, unlike Pearson's correlation coefficient. This allows \Dcorr to serve as the test statistic for an independence test. Finally, recent statistical results have shown that distance- and kernel-based methods are equivalent, allowing any independence test to be reformulated as a $k$-sample test for equality in distribution \cite{shenExactEquivalenceDistance2020}.

\begin{figure}[!t]
    \includegraphics[width=\linewidth]{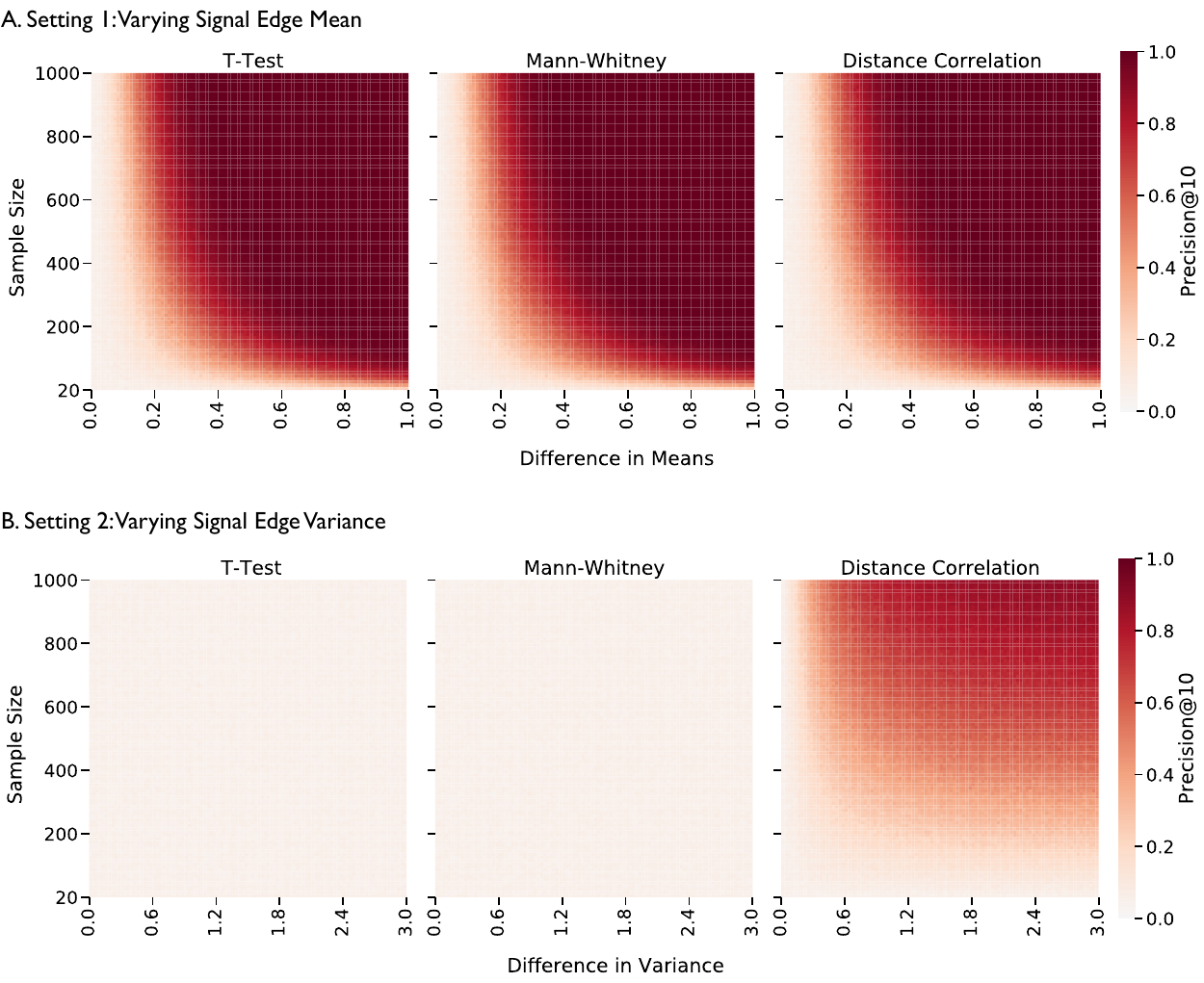}
    \caption{Precision@10 for each edge when comparing two populations of weighted networks using the \ttest, Mann-Whitney U test, and $k$-sample \Dcorr test. The color bar represents precision averaged over 100 trials. 
    \textit{(Top row)} Results for varying the mean $\delta$ and sample size while keeping the variance constant ($\phi = 0$). In this setting, all three tests perform similarly, and can detect signal edges when edge distributions differ in means.
    \textit{(Bottom row)} Results for varying the variance $\phi$ and sample size while keeping the mean constant ($\delta = 0$). \ttest and Mann-Whitney test cannot detect changes in variance regardless of the sample and effect size. $k$-sample \Dcorr test is the only test that can detect signal edges with changes in variance.}
    \label{fig:edgesim}
\end{figure}

\paragraph{Simulation study.} 
To illustrate the advantages of \Dcorr, we consider a two-population simulation setting where edge weights are sampled from distinct truncated normal distributions. When all signal edges are different only in their mean, \Dcorr, \anova, and Kruskal--Wallis all successfully identify signal edges, and no test outperforms the others in this setting (Figure \ref{fig:edgesim}A); however, when the mean edge weight is held equal but the variance is different across groups, only \Dcorr successfully detects the signal edges (Figure \ref{fig:edgesim}B). \anova and Kruskal--Wallis fail in this scenario because both test for differences in \textit{location} (the mean and median, respectively) across groups, whereas \Dcorr tests for differences in \textit{distribution}. This is of particular concern because many connectomics methods which aggregate serial edge-wise tests (e.g., the network-based statistic and its various extensions \cite{zaleskyNetworkbasedStatisticIdentifying2010, baggioStatisticalInferenceBrain2018, yooDegreeBasedStatistic2017}) use \anova to detect signal edges, and this theoretical result demonstrates a scenario in which these methods produce a spurious result. For all implementation details, see the Simulations (\S\ref{sec:edgesim}).

\paragraph{Real data experiments.} 
Using our edge-wise testing framework, we identified the signal edges across our four mouse strains. Application of \Dcorr showed that there was insufficient evidence to reject the null hypothesis of unequal edge weight distributions for any edge at $\alpha = 0.05$ following Holm--Bonferroni correction (for more information on how we corrected for multiple hypothesis testing, see the Methods (\S\ref{sec:correction})). However, this is more-so a result of sample size and the stringency of family-wise error rate (FWER) control than it is a result of lack-of-signal: when we visualize the tractogram and edge weight distribution for the edge with the smallest p-value (left hemisphere corpus callosum to right hemisphere striatum), we can appreciate clear differences. For example, the edge is much sparser in the BTBR mouse and the spatial distribution of the edge in the mouse brain is highly heterogenous across strain (Figure \ref{fig:main}). Therefore, to identify the \textit{strongest} signal edges (i.e., the connections between brain regions whose wiring patterns were most heterogeneous across different genotypes), we use the ranking of the p-values for each edge. Note that choosing to rank either the p-values or the associated test statistics for a set of edges will produce the same outcome: the edges that are most connectively distinct across different groups will be top ranked in any of these metrics. For example, the edge with the smallest p-value (i.e., strongest signal edge) will also have the largest test statistic. Both of these metrics effectively act as pseudo-dissimilarity measures, quantifying the joint distance between an edge across groups. 

\begin{table*}[!ht]
\centering
\begin{tabular}{@{}llll@{}}
\toprule
\textbf{Vertex 1}                   & \textbf{Vertex 2}                  & \textbf{statistic} & \textbf{p-value} \\ \midrule
Corpus Callosum (L)                 & Striatum (R)                       & 0.717              & 0.054            \\
Corpus Callosum (L)                 & Internal Capsule (R)               & 0.699              & 0.073            \\
Corpus Callosum (L)                 & Reticular Nucleus of Thalamus (R)  & 0.698              & 0.074            \\
Corpus Callosum (L)                 & Zona Incerta (R)                   & 0.686              & 0.092            \\
Septum (R)                          & Corpus Callosum (R)                & 0.671              & 0.118            \\
Lateral Ventricle (L)               & Striatum (R)                       & 0.667              & 0.125            \\
Striatum (L)                        & Striatum (R)                       & 0.664              & 0.132            \\
Corpus Callosum (L)                 & Ventral Thalamic Nuclei (R)        & 0.663              & 0.133            \\
Hippocampus (L)                     & Middle Cerebellar Peduncle (L)     & 0.658              & 0.144            \\
Caudomedial Entorhinal Cortex (R)   & Ventral Hippocampal Commissure (R) & 0.656              & 0.150            \\
Corpus Callosum (L)                 & Midbrain Reticular Nucleus (R)     & 0.653              & 0.159            \\
Midbrain Reticular Nucleus (L)      & Superior Cerebellar Peduncle (L)   & 0.648              & 0.172            \\
Corpus Callosum (L)                 & Corpus Callosum (R)                & 0.646              & 0.179            \\
Spinal Trigeminal Nerve (L)         & Middle Cerebellar Peduncle (L)     & 0.645              & 0.180            \\
Secondary Visual Cortex L (L)       & Striatum (R)                       & 0.641              & 0.192            \\
Globus Pallidus (R)                 & Midbrain Reticular Nucleus (R)     & 0.632              & 0.225            \\
Striatum (L)                        & Corpus Callosum (R)                & 0.632              & 0.226            \\
Primary Somatosensory Cortex HL (L) & Secondary Visual Cortex MM (L)     & 0.629              & 0.236            \\
Corpus Callosum (L)                 & Primary Somatosensory Cortex J (R) & 0.628              & 0.238            \\
Corpus Callosum (L)                 & Ventral Orbital Cortex (R)         & 0.628              & 0.240            \\ \bottomrule
\end{tabular}
\caption{The top 20 signal edges (out of 52,975 total edges) ranked by the order of their Holm--Bonferroni corrected p-value. Eleven of the top 20 signal edges are adjacent to either the left or right hemisphere corpus callosum.}
\label{tab:edge}
\end{table*}

A list of the 20 strongest signal edges, along with their corresponding \Dcorr test statistics and p-values, are given in Table \ref{tab:edge}. The strongest signal edge connects the left hemisphere corpus callosum to the right hemisphere striatum. In fact, 13 of the 20 strongest signal edges are incident to these two ROIs, demonstrating that the connections emanating from these regions are highly heterogeneous across genotypes. Since connective abnormalities in edges are highly correlated with specific regions of the brain, this finding suggests that in addition to identifying signal \textit{edges}, it is also insightful to interrogate the next hierarchical level of network topology and identify signal \textit{vertices}. 

\subsubsection{Identifying signal vertices.}
The ability to discover brain regions that are topologically dissimilar across phenotypes (i.e., \textit{signal vertices}) is critical for scientific and clinical analyses of connectomes, with broad applications such as the establishment of neurological biomarkers and identification of therapeutic targets \cite{craddockConnectomicsNewApproaches2015a}. Here, we leverage recent advances in the theory of random graph models to propose a principled and robust statistical method for identifying signal vertices. We also demonstrate that this method recovers more information about ROIs than vertex-level graph statistics, the predominant method for analyzing vertices in multi-subject connectomics \cite{craddockImagingHumanConnectomes2013a, fornitoGraphAnalysisHuman2013, wangGRETNAGraphTheoretical2015}. Specifically, we use the omnibus embedding (\omni) \cite{athreyaStatisticalInferenceRandom2017a}---a vertex embedding technique---to jointly represent all the connectomes in a multi-subject dataset within a common Euclidean subspace. This latent position quantifies a given vertex's probability of connecting to any other vertex in the connectome (for more details, see the Methods (\S\ref{sec:models})). For each vertex, we then apply a multivariate statistical test to determine whether the embedding of a specific brain region is different across groups. Previous results have shown that the latent position vectors estimated by \omni are asymptotically Gaussian \cite{levinCentralLimitTheorem2017b}, motivating our use of multivariate analysis of variance (\manova) to identify the signal vertices.

\paragraph{Simulation study.} 
To illustrate the advantages of this approach over existing techniques, we compare \omni to three other previously established vertex embedding methods: (i) the Exponential random graph model (ERGM), which incorporates the approach of representing each vertex with a vector of graph statistics \cite{simpsonExponentialRandomGraph2011}; (ii) Multivariate Distance Matrix Regression (MDMR), which represents each vertex as a vector of the weights for all adjacent edges \cite{kimComparisonStatisticalTests2014}; and (iii) the Network-Based Statistic (NBS), which identifies ROIs in subgraphs comprised exclusively of strong signal edges as signal vertices \cite{zaleskyNetworkbasedStatisticIdentifying2010}. In a two-population simulation setting, we sample graphs from a distribution in which the true number of signal vertices is determined \textit{a priori}. Then, for each vertex, we compute a p-value using each of the three vertex embeddings followed by Holm--Bonferroni correction. We measure the statistical power of each approach via a Receiver Operating Characteristic (ROC) curve, a performance metric which holistically describes the performance of a binary classifier by characterizing the trade-off between sensitivity and specificity as the discriminant threshold of the classifier (in this case, $\alpha$) is varied. For all implementation details, see the Simulations (\S \ref{sec:nodesim}).

\begin{figure*}[t]
    \centering
    \includegraphics[width=\linewidth]{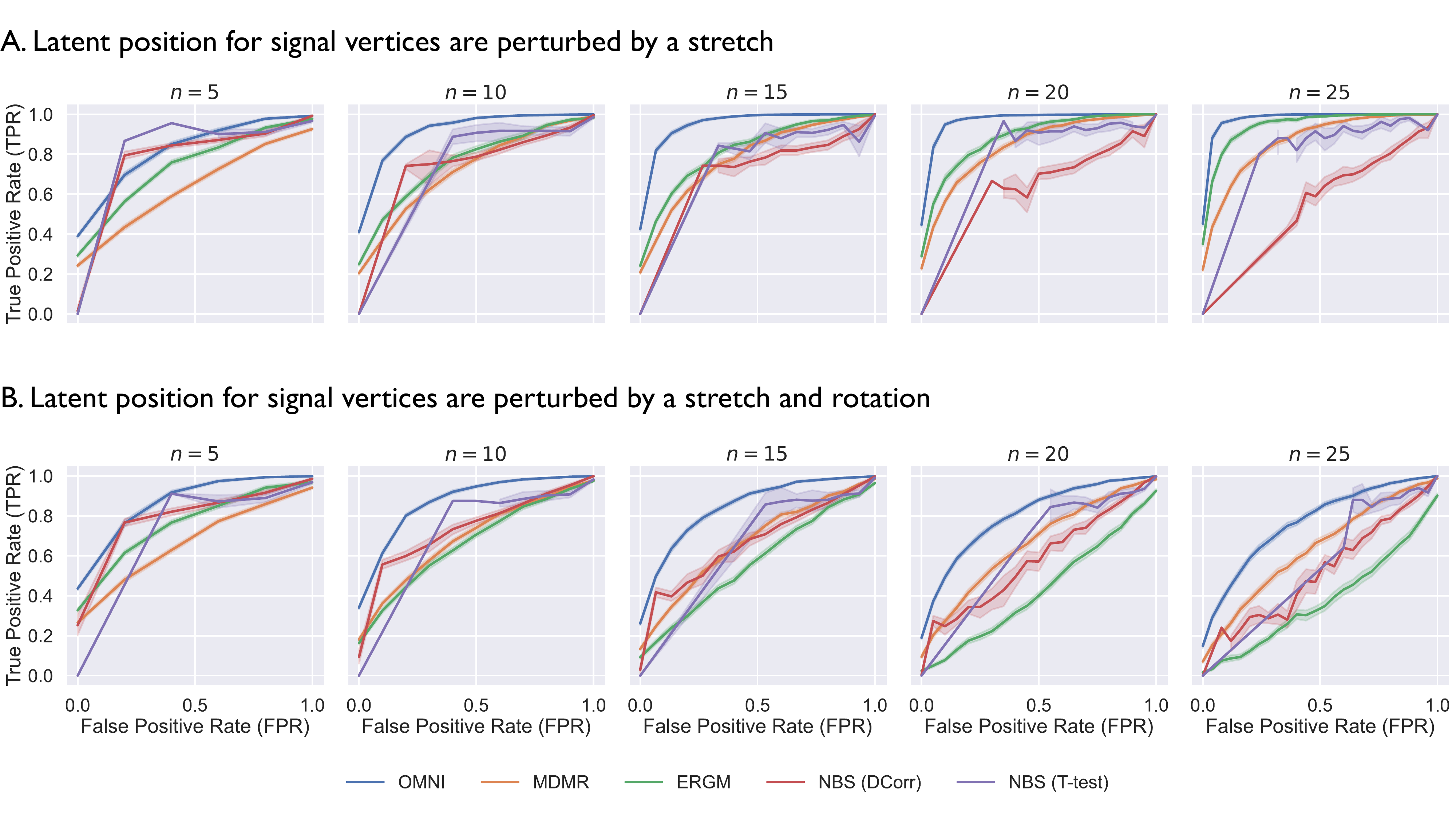}
    \caption{Ability of various vertex representations to identify signal vertices. The number of vertices in each network is kept constant ($50$), but the number of signal vertices is changed ($n = 5, 10, 15, 20, 25$). For each vertex, we compute a p-value from each of the three vertex representations using \hotellings, and set the significance level at $\alpha=0.05$ following Holm--Bonferroni correction. Colors correspond to the method of vertex representation. \textit{(Top row)} This setting compares two different \rdpg s where the perturbed latent position is stretched by a constant. \textit{(Bottom row)} This setting compares \rdpg s where the perturbed latent position is rotated and stretched by a constant. Both settings show that \omni successfully differentiates null and signal vertices (AUC=87\%) with much more accuracy than MDMR (AUC=73\%), NBS (AUC=59\% with the \ttest), or the ERGM (AUC=68\%). Replacing the \ttest in NBS with \Dcorr marginally improved the AUC to 62\%, but the performance of this method is still far worse than than \omni.}
    \label{fig:nodesim}
\end{figure*}

Figure \ref{fig:nodesim} shows that \omni is a superior vertex representation method to the ERGM, MDMR, or NBS in this simulation, with an average Area Under the ROC (AUROC) 20\% greater than the other three methods. Even in the most challenging simulation settings we test, where the true number of signal vertices is very small, \omni correctly identifies the vertices of interest with much higher accuracy than the other three methods. The ERGM in particular performed very poorly, achieving accuracy on par with random change in the most challenging settings (Figure \ref{fig:nodesim} \textit{bottom right panel}). We hypothesize that the poor performance of the ERGM results from the arbitrary choice of graph statistics used to parameterize the model. As Simpson et al. note in their original ERGM paper, ``the most appropriate explanatory metrics vary by network'' \cite{simpsonExponentialRandomGraph2011}. Thus, choosing an appropriate set of graph statistics for a given multi-subject connectomics dataset is a highly subjective task, limiting reproducibility of this method across studies. While a given graph statistic might be insightful given \textit{a priori} knowledge about the data one is analyzing, they are not the most informative vertex embedding for any generic multi-subject connectomics dataset. In contrast, the \omni-based test we propose is a dataset-agnostic prescriptive statistical procedure that does not require hand tuning.

\paragraph{Real data experiments.} 
As it has been previously reported that the BTBR mouse has significant neuroanatomical abnormalities in the corpus callosum \cite{meyzaBTBRMouseModel2017, ellegoodNeuroanatomicalAnalysisBTBR2013a}, we expect our method to identify this brain region as a strong signal vertex. Our results corroborate this hypothesis: across all mouse strains, the left hemispheric portion of the corpus callosum is the strongest signal vertex, and its right hemispheric counterpart is the second strongest (note, because this atlas is symmetric, the portions of the corpus callosum in each hemisphere are parcellated as two distinct ROIs) (Table \ref{tab:vertex}). In contrast, existing connectomics methods like MDMR fail to identify this salient neurobioligical information: as shown in Table \ref{tab:mdmr_vertex}, MDMR ranks the left corpus callosum as the 34th strongest signal vertex despite its pronounced connective differences across strains. We also attempt to use vertex-level graph statistics (degree, clustering coefficient, betweenness centrality, closeness centrality, and number of triangles) to identify the connective abnormalities of the corpus callosum: however, the corpus callosum was the 92nd strongest signal vertex using this method. Like other existing embedding techniques not based on random graph models, MDMR and vertex-level graph statistics do not enjoy the statistical advantages of \omni, including its interpretability and theoretical foundations.

\begin{figure*}[t]
    \includegraphics[width=\linewidth]{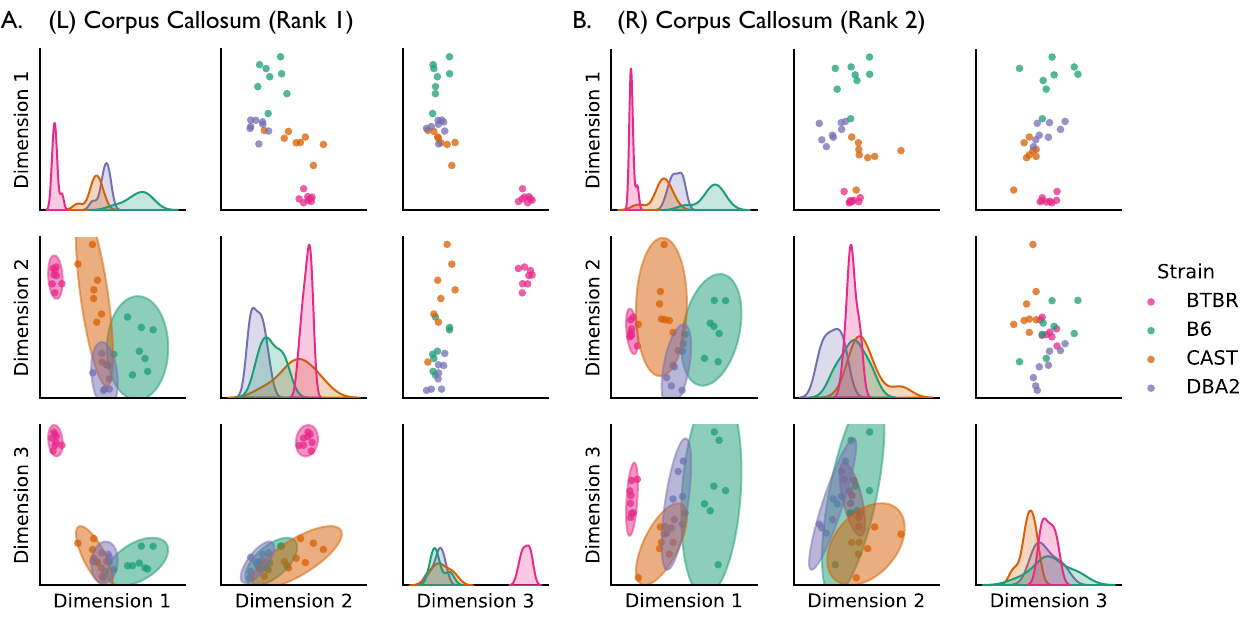}
    \caption{Pairs plots of vertex embedding of the left and right hemisphere corpus callosum produced by the omnibus embedding. The corpus callosum is the bridge between the left and right hemispheres of the brain. In the BTBR mouse, the corpus callosum has highly penetrant neuroanatomical defects; therefore, we expect to differentiate the BTBR mouse from the other strains by examining the embedding of the corpus callosum. We can do this in the embedding of the left hemisphere corpus callosum (the strongest signal vertex) and slightly less clearly in the embedding for the right hemisphere corpus callosum (the second strongest signal vertex). Interestingly, corpus callosum embeddings of wild-type mice are also separable (particularly the B6 strain), suggesting a diversity in corpus callosum architecture across the behavioral controls.}
\label{fig:corpus}
\end{figure*}

\begin{table}[ht!]
\centering
\begin{tabular}{@{}llll@{}}
\toprule
\textbf{Vertex}                & \textbf{Pillai} & \textbf{F} & \textbf{p-value} \\ \midrule
Corpus Callosum (L)            & 2.591           & 32.91             & 5.09e-25         \\
Corpus Callosum (R)            & 2.556           & 29.95             & 1.09e-23         \\
Fimbria (L)                    & 2.440           & 22.64             & 7.27e-20         \\
Secondary Motor Cortex (L)     & 2.438           & 22.54             & 8.21e-20         \\
Midbrain Reticular Nucleus (R) & 2.430           & 22.16             & 1.38e-19         \\
Substantia Nigra (R)           & 2.305           & 17.25             & 2.20e-16         \\
Internal Capsule (R)           & 2.304           & 17.23             & 2.29e-16         \\
Secondary Motor Cortex (R)     & 2.297           & 16.99             & 3.40e-16         \\
Cerebral Peduncle (R)          & 2.247           & 15.51             & 4.34e-15         \\
Internal Capsule (L)           & 2.238           & 15.27             & 6.71e-15         \\
Striatum (L)                   & 2.236           & 15.23             & 7.13e-15         \\
Lateral Ventricle (L)          & 2.218           & 14.74             & 1.74e-14         \\
Stria Terminalis (R)           & 2.202           & 14.35             & 3.59e-14         \\
Cerebellar White Matter (R)    & 2.199           & 14.28             & 4.08e-14         \\
Optic Tracts (L)               & 2.186           & 13.96             & 7.52e-14         \\
Subthalamic Nucleus (L)        & 2.178           & 13.78             & 1.05e-13         \\
Hippocampus (R)                & 2.177           & 13.76             & 1.08e-13         \\
Stria Terminalis (L)           & 2.177           & 13.75             & 1.11e-13         \\
Frontal Association Cortex (L) & 2.170           & 13.60             & 1.47e-13         \\
Rostral Linear Nucleus (R)     & 2.165           & 13.47             & 1.88e-13         \\ \bottomrule
\end{tabular}
\caption{The top 20 signal vertices (out of 326 total vertices) ranked by the order of their Holm--Bonferroni corrected p-values. Pillai's trace and approximate $F$ statistic (along with 15,78 degrees of freedom) as calculated by one-way \manova~ are also reported. The corpus callosum in the left and right hemisphere are the top two signal vertices.}
\label{tab:vertex}
\end{table}

In Figure \ref{fig:corpus}, we plot the vertex embedding of the corpus callosum obtained by \omni using a pairs plot \cite{emersonGeneralizedPairsPlot2013}. Because \omni embeds each vertex of the graph in $d$-dimensional space, the pairs plot allows us to visualize the high-dimensional relationships in this Euclidean representation of network connectivity. Each dot represents the embedded corpus callosum of an individual mouse. These plots show scatter plot matrices on the off-diagonal panels, with kernel density estimates (KDEs) of the marginal distributions (smooth approximations of the underlying distribution of the data) on the diagonal. In the lower triangle of each pairs plot, we show 95\% prediction intervals for each strain's vertex embedding. Together with the KDEs, these figures show a high degree of separability in the embeddings, highlighting the intra-strain heterogeneity of the corpus callosum.  Thus, \omni successfully recovers a distinct representation of the corpus callosum in BTBR mice, and the corroboration of known neurobiological results adds further validation to this approach. For comparison, in Supplementary Figure \ref{fig:weak_vertex}, we also show pairs plots of two weak signal vertices: the left hemisphere cingulate cortex area 29c (rank 170 of 326) and the right hemisphere fasciculus retroflexus (rank 257 of 326). Mice from distinct genotypes are much less separable in embeddings of these vertices compared to the embeddings of the corpus callosum.

Connectomes in this dataset are bilateral (that is, for each ROI, there is a corresponding structure in both the left and right hemispheres). Therefore, we can aggregate the connective abnormality of a structure across hemispheres, enabling the identification of pairs of vertices that are significant in both the left and right hemisphere. In Table \ref{tab:bilateral-vertices}, we provide a list of the 10 strongest signal vertex pairs, which we term bilateral signal vertices. For most bilateral signal vertices, the ROI in one hemisphere is usually a much stronger signal vertex than the other hemisphere (for example, the right hemisphere cerebral peduncle is the 9th ranked signal vertex, while its right hemisphere counterpart is ranked 63rd).

\subsection{Interrogating the \textit{regional scale}.}
\label{sec:meso}
Communities of highly interconnected vertices are important structures within connectomes that underlie diverse neurological functions \cite{betzelDiversityMesoscaleArchitecture2018, vandenheuvelComparativeConnectomics2016a}. Therefore, communities are an important topological level at which connectomes can be analyzed. 

\subsubsection{Identifying signal communities.}
For analysis of connectomes where vertices are organized using an \textit{a priori} community grouping, we compare four approaches for modelling the connectivity information encoded within a community (see the Methods (\S \ref{sec:regional-detail})). Each successive approach yields a more holistic summary of the community. The first two approaches are univariate, comparing either (1) the probability of connectivity in a community or (2) the average edge weight in a community across populations. While these are fundamental properties of a community, summarizing the behavior of a community with a single scalar loses information. The last two approaches are multivariate, comparing either (3) the indices of nonzero edges in a community or (4) the vector of edge weights in a community. Note that the operations performed by approaches (1) and (3) require the connectomes to be binarized via Otsu's method \cite{otsuThresholdSelectionMethod1979}, whereas approaches (2) and (4) operate on weighted connectomes. To test if the summarized information in a block is different across phenotypes, we again use \Dcorr.

\begin{figure}[!b]
    \centering
    \includegraphics[width=\linewidth]{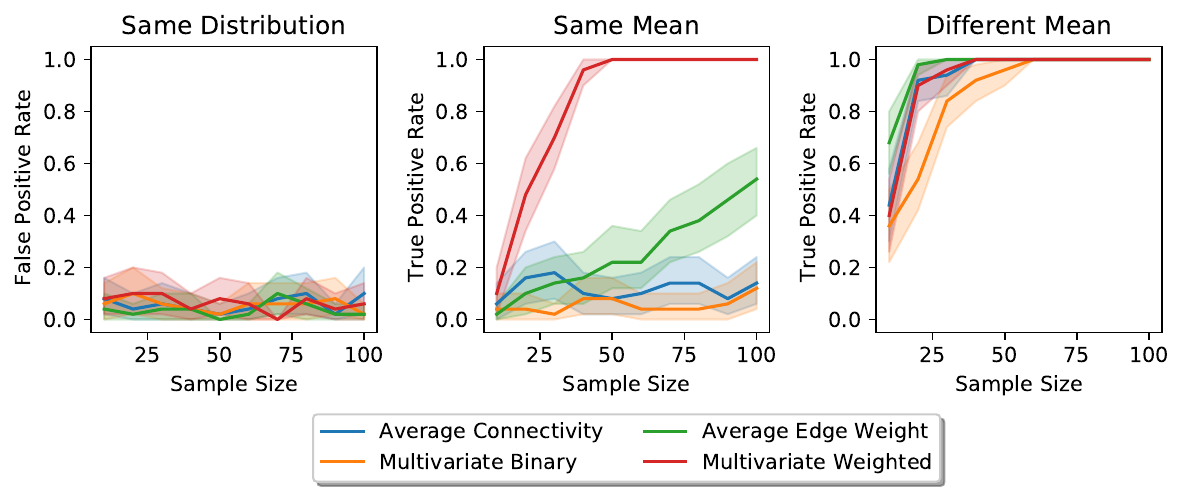}
    \caption{The Multivariate Weighted approach for signal community detection is superior to other proposed methods. We consider a two-population simulation where edge weights are sampled from 3-community block diagonal \sbm. Distributions in the first block are equal, allowing us to measure the False Positive Rate for each method. We see that all methods achieve a FPR less than or equal to $\alpha$ (\textit{left}). In the second block, the two distributions have the same mean, but different variances. The Multivariate Weighted method is the most able to detect community edges in this setting, however it requires a sample size of at least $N=25$ (\textit{middle}). In the third block, the two distributions have different means, and in this setting, the Average Edge Weight method is superior given small sample sizes ($N<25$) (\textit{right}). However, when the samples size is sufficiently large, the Multivariate Weighted method is equivalent. These results demonstrate the superiority of the Multivariate Weighted method over other proposed methods.}
    \label{fig:commsim}
\end{figure}

\paragraph{Simulation study} We compare these approaches in a two-population simulation setting (see the Simulations (\S \ref{sec:commsim})). All methods were robust to false positives (Figure \ref{fig:commsim} \textit{first panel}), however, they differed in their ability to successfully identify signal communities. In settings where edge weights in a community have the same mean but different variances, univariate and binary approaches struggle to identify signal communities, achieving a maximum TPR of 60\%. However, comparing multivariate weighted representations of communities proved much more successful with a TPR of 80\% for sample sizes $N > 30$ (Figure \ref{fig:commsim} \textit{middle panel}). When edge weights in a community have different means, all algorithms are able to successfully identify signal communities with a high TPR. However, only the weighted approaches can do this with small sample sizes $N < 25$ (Figure \ref{fig:commsim} \textit{right panel}). Therefore, we propose using \Dcorr to compare communities across subjects in \textit{regional level} analyses.

\begin{figure*}[hbtp]
\centering
\includegraphics[width=0.765\linewidth]{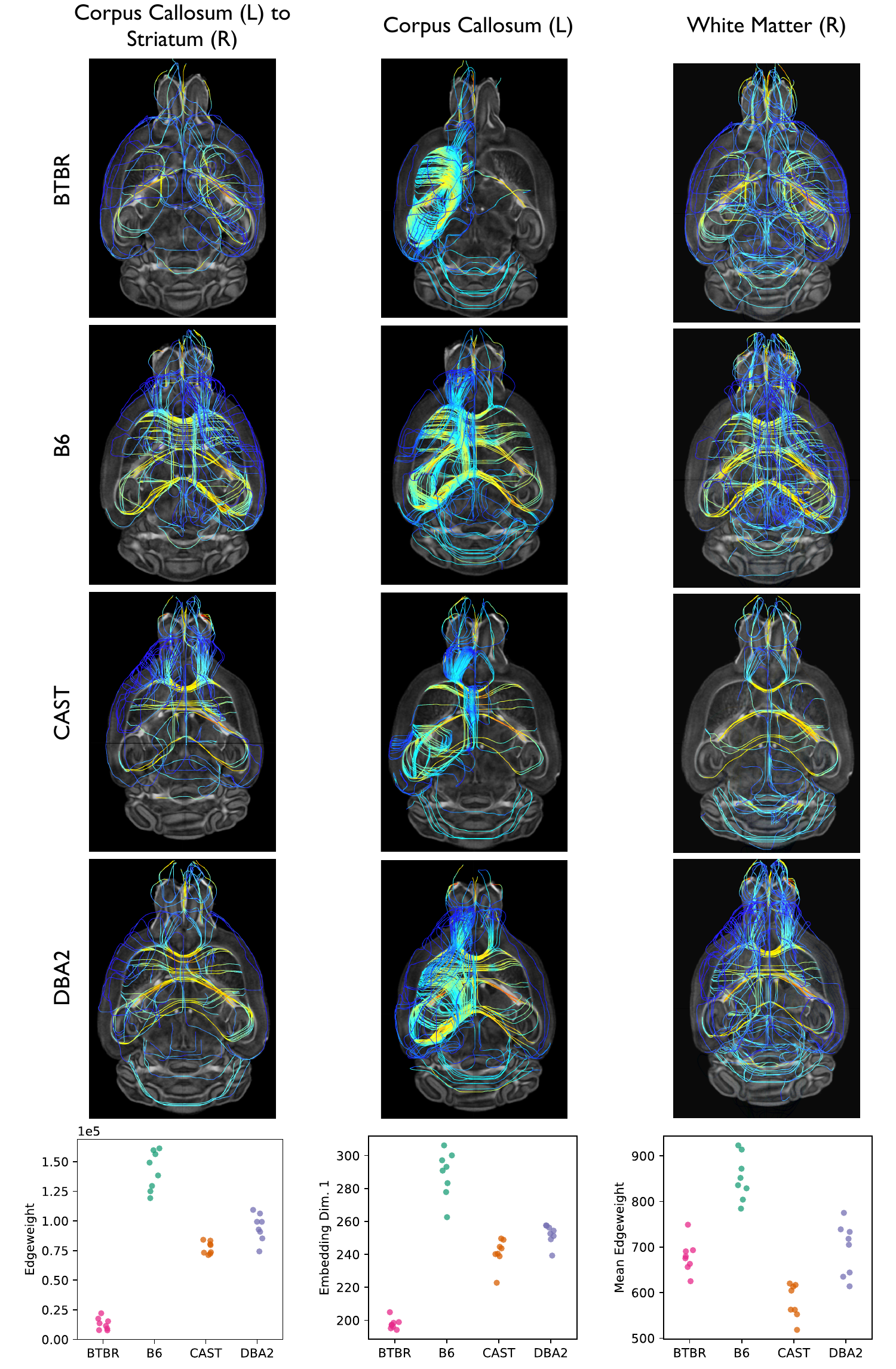}
    \caption{Visualization of the strongest signal edge (left hemisphere corpus callosum to right hemisphere striatum), vertex (left hemisphere corpus callosum), and community (right hemisphere white matter) across all mouse strains. At each topological level, tractograms of these neurological structures are shown for each mouse strain. (\textit{Bottom row}) The distribution of edge weights for the strongest signal edge (\textit{Column 1}); the distribution of the first embedding dimension for the strongest signal vertex (\textit{Column 2}); and the distribution of average edge weight for the strongest signal community (\textit{Column 3}). Each dot represents data from an individual mouse. Connective differences in the left hemisphere corpus callosum are apparent, with BTBR mice displaying a uniformly small vertex embedding. The average edge weight of the most significant community also shows pronounced variation across strains.}
\label{fig:main}
\end{figure*}

\paragraph{Real data experiments} In Supplementary Figure \ref{fig:mgc}, we show log-transformed $p$-values obtained by the four approaches described above. Regions in blue are significant at the Holm--Bonferroni correction.  Consistent with the simulations in the Simulations (\S \ref{sec:commsim}), we see that univariate tests find less signal communities than the multivariate tests. $p$-values for all communities are given in Table \ref{tab:communities}. The strongest signal community (i.e., the one with the most heterogeneous topology across genotypes) determined by \Dcorr is the intraconnection within the right hemisphere white matter. In fact, the majority of the 10 strongest signal communities involve connections to the white matter in both hemispheres.

\begin{table}[t]
\centering
\begin{tabular}{@{}llll@{}}
\toprule
\textbf{Community 1} & \textbf{Community 2} & \textbf{statistic} & \textbf{p-value} \\ \midrule
White Matter (R) & White Matter (R) & 0.885 & 6.43e-06 \\
White Matter (L) & White Matter (L) & 0.857 & 1.02e-05 \\ 
Hindbrain (L) & White Matter (L) & 0.849 & 1.14e-05 \\
Midbrain (R) & White Matter (R) & 0.845 & 1.21e-05 \\
Isocortex (L) & Isocortex (L) & 0.844 & 1.22e-05 \\
Pallium (R) & White Matter (R) & 0.831 & 1.5e-05 \\
Isocortex (R) & White Matter (R) & 0.823 & 1.68e-05 \\
Isocortex (R) & Isocortex (R) & 0.819 & 1.8e-05 \\
Isocortex (L) & White Matter (L) & 0.811 & 2.02e-05 \\
Hindbrain (R) & White Matter (R) & 0.810 & 2.02e-05 \\ \bottomrule
\end{tabular}
\caption{The top 10 signal communities (out of 105 total communities) ranked by the order of their Holm--Bonferroni corrected p-values as calculated by the Multivariate Weighted method.}
\label{tab:communities}
\end{table}

\begin{figure*}[t]
\centering
\includegraphics[width=\linewidth]{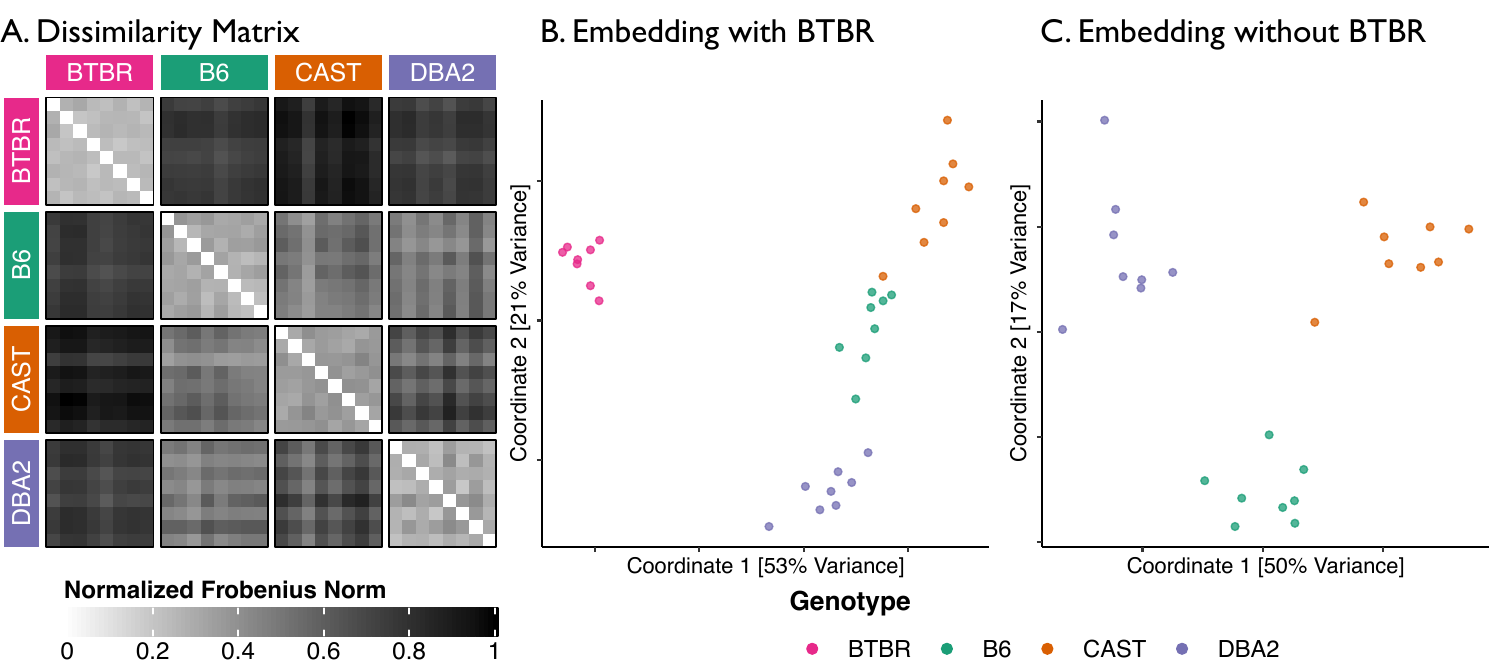}
    \caption{Pairwise dissimilarity between each mouse connectome, organized by mouse strain (\textit{left}) and the joint embeddings of each sample with and without connectomes from BTBR mice in a two-dimensional space (\textit{right} and \textit{center}, respectively). Joint embeddings of every connectome were obtained using the omnibus embedding. (\textit{Left}) The pairwise dissimilarity between connectomes is calculated as the Frobenius norm of the difference between the embeddings of a pair of connectomes. (\textit{Center}) Two-dimensional representations of each connectome were obtained by using Classical Multidimensional Scaling (\cmds) to reduce the dimensionality of the embeddings obtained by \omni. (\textit{Right}) Same as center, but without data from BTBR mice.}
\label{fig:cmds}
\end{figure*}

\subsection{Identifying multiscale differences in network architecture across genotypes.}
Using the statistical methods described above, we discover differences in brain connectivity between the four mouse genotypes at each topological level of the connectome. Figure \ref{fig:main} visualizes the strongest signal edge (as detected by \Dcorr), the strongest signal vertex (as detected by \omni and \manova), and the strongest signal community (as detected by \Dcorr) using tractograms, renderings of nerve tracts measured in the original DTI data. For an edge, the tractogram represents all the tracts between its two incident vertices; for a vertex, the tractogram represents all the tracts originating from that vertex; and for a community, the tractogram represents all the tracts interconnecting the vertices in a superstructure. A full list of the parameters using to generate these tractograms is available in the Methods (\S\ref{sec:tractography}).

Tractograms allow us to visualize the heterogeneity in brain connectivity identified by our multi-level algorithms. For example, a tractogram of the left corpus callosum (Figure \ref{fig:main} \textit{middle column}) in the BTBR mice reveals a near absence of cross-hemispheric connections, while all control strains display much more cross-hemispheric connections at this vertex. To contrast with these heterogeneous tractograms, Supplementary Figure \ref{fig:sup-main} shows tractograms for the weakest signal edge, vertex, and community. In comparison to the tractograms shown in Figure \ref{fig:main}, the tractograms of weak signal components are much more homogeneous across strains. Note, in the calculation of the weakest signal edge, we ignore edges with zero edge weight for all connectomes. 

In addition to tractograms, we also plot the distribution of neurotopological features (edge weights and embeddings) used to determine the strongest signal structure at each level (Figure \ref{fig:main} \textit{bottom row}). These distributions highlight the differences in connectome structure that each algorithm used to quantify the signal strength of that specific edge, vertex, or community.

\begin{figure*}[t]
\centering
\includegraphics[width=0.8\linewidth]{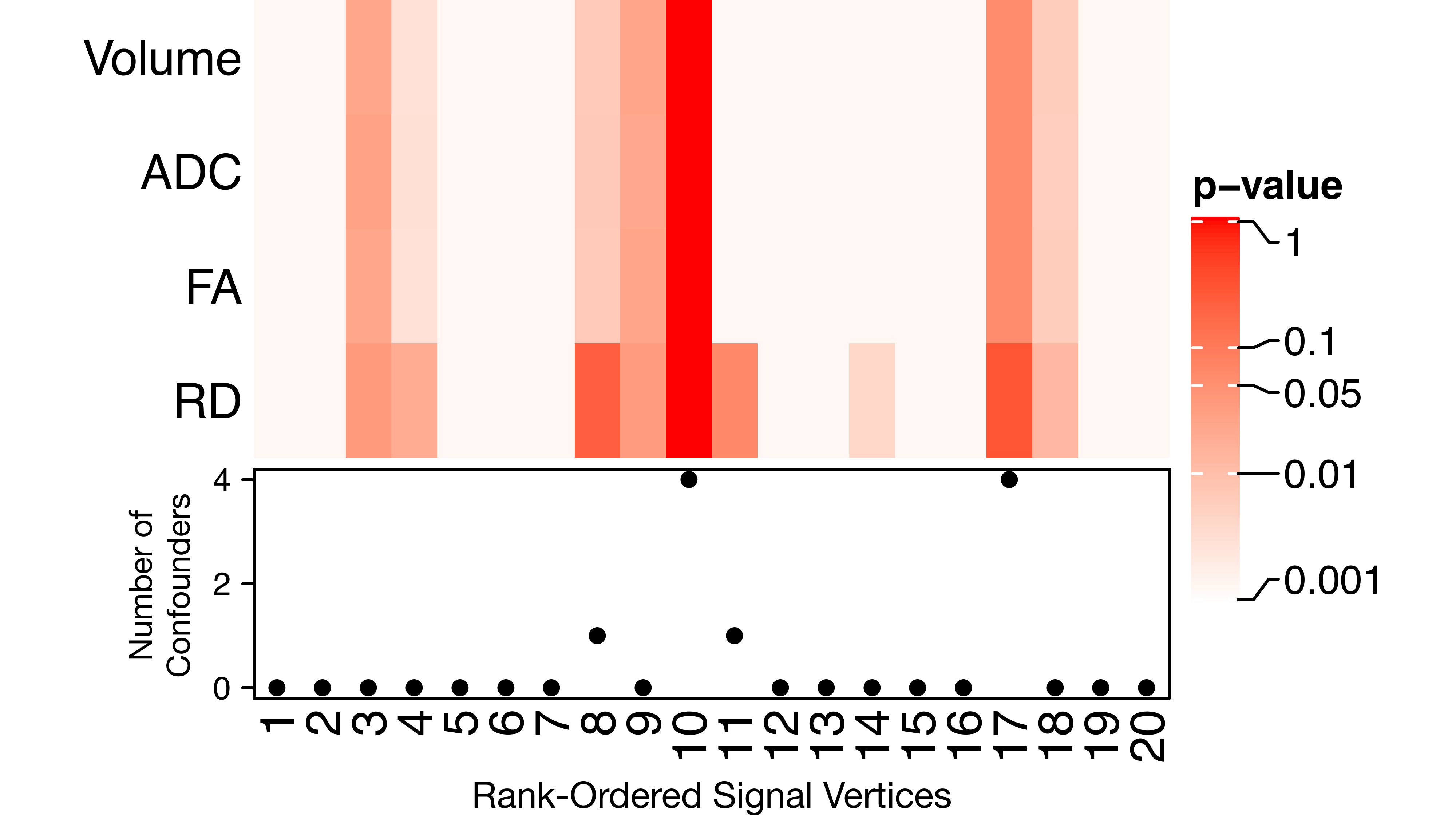}
    \caption{The omnibus embedding provides a novel understanding of brain connectivity at the level of individual vertices. Using diffusion imaging, the following metrics were measured for each vertex: volume, apparent diffusion coefficient (ADC), fractional anisotropy (FA), and radial diffusivity (RD). To determine if a given feature provided more information than the embedding, these four anatomical features were compared to vertex embeddings produced by \omni via a conditional independence test. The Holm--Bonferroni corrected p-values for conditional independence tests for each of the top 20 strongest signal vertices (Table \ref{tab:vertex}) are shown on the heatmap above. A cell is red if genotype and the embedding are conditionally independent given the feature, implying that \omni is confounded by the anatomical feature. The total number of confounding anatomical features for each vertex, sorted by signal strength are shown below the heatmap.}
\label{fig:causal}
\end{figure*}

\subsection{Interrogating the \textit{global scale}.}
To complete a multi-level analysis of multi-subject connectomics data, we demonstrate how results from our prior levels of network topology can be aggregated to enable comparisons of patterns in whole-brain connectivity across subjects. Figure \ref{fig:cmds}A shows the dissimilarity between each pair of connectomes in our dataset, where pairwise dissimilarity between connectomes $G_1$ and $G_2$ is calculated as the Frobenius norm of the difference between the corresponding embedding of each connectome calculated by \omni. Note that the dissimilarities in Figure \ref{fig:cmds}A are standardized by dividing by the largest pairwise dissimilarity. The average intra-strain dissimilarity (27\%) is smaller than the average inter-strain dissimilarity (67\%), confirming that connectomes from mice of the same genotype are globally most similar to one another. Additionally, the BTBR mice are very dissimilar to all other strains (the average inter-strain dissimilarity for the BTBR mice is 79\%) while the three control strains are all fairly similar to each other (the average inter-strain dissimilarity for B6, CAST, and DBA2 mice is 55\%). 

We further reduce the dimensionality of these embeddings by using classical Multidimensional Scaling (\cmds) to embed this dissimilarity matrix into a two-dimensional space \cite{coxMultidimensionalScaling2008}. This yields a collection of 32 points in $\mathbb{R}^2 \,,$ where each point represents the connectome of an individual mouse. The BTBR mice are highly separated from the other three control strains (Figure \ref{fig:cmds}B); all wild-type strains are also clearly distinct from each other (Figure \ref{fig:cmds}C). Thus, we can leverage information from \omni to successfully differentiate all connectomes based on genotype, enabling comparisons of brain connectivity at the \textit{global level}.

\subsection{The topology of vertices encodes information beyond the anatomy of those vertices.}
Here, we demonstrate that the characterization of vertex-level brain connectivity provided by the omnibus embedding contains topological information that is not available in commonly-used anatomical features. Following registration of all diffusion imaging data, the following anatomical features were derived for each vertex in every mouse brain: volume, apparent diffusion coefficient (ADC), fractional anisotropy (FA), and radial diffusivity (RD). For every vertex, we tested if the genotype labels and the low-dimensional embedding of the vertex produced by \omni were conditionally independent given any of the anatomical features of that vertex \cite{wangSureIndependenceScreening2017a}. Phrasing this mathematically, let $Y \in \mathcal{Y}$ be a discrete random variable representing the genotype of a mouse. For a given vertex $i$, let $X_i \in \mathcal{X} \subseteq \mathbb R^d$ be a random variable representing the vertex's latent position in an \rdpg estimated by \omni, and let $A_i \in \mathcal{A}$ be a vector of anatomical features that describe the vertex. If the mouse's genotype and the vertex's embedding are conditionally independent given the anatomical features, then
\begin{equation}
    (Y \pperp X_i) \mid A_i \implies \Pr(Y \mid X_i, A_i) = \Pr(Y \mid A_i) \,.
\end{equation}
That is, the information about connectivity encoded in the vertex's latent position is redundant given the anatomical features. However, if null hypothesis of conditional independence is rejected, the latent position contains information about connectivity not represented in the anatomy.

We find that for 282 out of 326 vertices, the genotype labels and \omni embedding were conditionally dependent given each of the anatomical feature (Figure \ref{fig:causal}). All four anatomical features were confounders for only 28 vertices, and the majority of these vertices were not in the top 100 strongest signal vertices. Thus, for most vertices, the omnibus embedding provides additional insight into connectivity beyond what is explained by the anatomical features, demonstrating the value of this means of \textit{local level} analysis. Our findings demonstrate that, in most cases, anatomical features do not encode additional information about mouse genotype than the connectomes.

\section{Discussion}
Robust and interpretable statistical methods for analyzing the network topology of connectomes are critically important to understand how patterns in brain connectivity are associated with observable neurological phenotypes. To this end, we leverage recent advances in the theory of random graph models to develop open-source methods that deliver statistically principled and interpretable analyses of multi-phenotype and multi-subject connectomics datasets. Specifically, these methods can be used to identify signal edges, vertices, and communities --- that is, the components of the connectome across multiple topological levels that characterize the differences in network architecture observed between samples from distinct phenotypic backgrounds. Additionally, we have formulated these methods as $k$-sample hypothesis tests, meaning they can be used to jointly analyze connectomes from more than just two dimensional or categorical phenotypes.

There are numerous advantages to using random graph models to perform hypothesis tests on connectomes:
\begin{enumerate}
    \item Hierarchical random graph models enable \textit{multi-level modeling} of connectomes; 
    \item Random graph models are generative, meaning the estimated model parameters characterize the structure of a particular level of network topology, and are therefore \textit{interpretable}; 
    \item Unlike many other connectomics methods, random graph models have \textit{provable properties} which motivate robust and principled downstream statistical analyses.
\end{enumerate}
For example, consider the identification of signal vertices using the Random Dot Product Graph (\rdpg) and \omni. A previously demonstrated central limit theorem proved that the embeddings estimated by \omni are asymptotically normal \cite{levinCentralLimitTheorem2017b}, motivating our use of \manova to test for connective differences amongst the ROIs in a connectomics dataset. In contrast, vertex-level graph statistics have no such governing distribution. In fact, multiple studies have shown that connectomes with wildly different topologies can produce identical graph statistics \cite{chenSameStatsDifferent2019, chungStatisticalConnectomics2020a} (\`a la Anscombe’s quartet \cite{anscombe_graphs_1973}), so while graph statistics can produce seemingly intuitive and logical summaries of connectome connectivity, these measures are generally unable to provide a holistic characterization of a vertex's connectivity. This is reflected in our simulation studies: successfully detecting signal vertices using vertex-level graph statistics is an under-powered statistical approach, with an average area under the ROC curve (AUROC) of 68\% (Supplementary Figure \ref{fig:nodesim}), lagging behind our proposed method, which achieves an average AUROC of 87\%.

We have previously proposed vertex embedding methods other than \omni, and while these methods have certain advantages over \omni in particular scenarios, they too lack many desirable statistical properties. For example, unlike \omni, the Joint Spectral Embedding \cite{nenningJointEmbeddingScalable2020} can embed connectomes with different numbers of vertices. While this is useful for cross-species comparisons \cite{xuCrossspeciesFunctionalAlignment2020}, this paradigm is not often encountered in multi-subject connectomics datasets of the same species (e.g., mice or humans) given that the same atlas can be applied to all subjects. One disadvantage of the Joint Spectral Embedding, however, is that its outputs are not interpretable. In \omni, the embedding of a given vertex (encoded as a vector) quantifies its probability of connecting to any other vertex in the connectome. In addition to lacking such interpretability, the Joint Spectral Embedding does not enjoy the asymptotic normality of \omni, limiting its application for identifying signal vertices.

Our methods for identifying signal edges, vertices, communities enable powerful multi-level analyses of connectomes. Additionally, we show how information from each of these levels can be aggregated to perform whole-brain comparisons of global level connectivity (Figure \ref{fig:cmds}). Previous methods have attempted to perform similar multi-level analyses, but, in addition to analyzing fewer levels than our methods are capable of, the theoretical foundations of these methods could be improved. For example, the network based statistic (NBS) aggregates the results from serial edgewise tests to find clusters of inter-connected vertices that are connectively different between two groups \cite{zaleskyNetworkbasedStatisticIdentifying2010}. However, NBS uses the \ttest and arbitrary thresholding to determine which edges in the connectome are signal edges. As we have shown in our simulations, the \ttest (as well as other tests of location, such as Kruskal--Wallis) can only successfully detect signal edges in well-behaved edge weight distributions (e.g., two normal distributions with equal variance and different means) (Supplementary Figure \ref{fig:edgesim}). Edge weights can be distributed differently across phenotypes, and therefore be true signal edges, but have the same mean (or median), and NBS will fail to successfully identify them as signal edges. To correct for this deficiency, we advocate for the use of \Dcorr, a nonparametric, universally consistent test for differences in distribution \cite{szekelyMeasuringTestingDependence2007, pandaHyppoComprehensiveMultivariate2020}. By treating each edge weight as an independent random variable under the Independent Edge (\ie) random graph model, we enable far more powerful edgewise testing. In the Simulations (\S\ref{sec:nodesim}), we substitute the \ttest in NBS with \Dcorr. While this modification improves the method's average AUC by about 3\%, both versions of NBS fail to robustly identify signal edges, and lag behind existing methods like MDMR and our \omni-based procedure. If one is truly interested in estimating vertex clusters like those produced by NBS, we instead recommend the signal subgraph estimator \cite{6341752, wangSignalSubgraphEstimation2018}, which has the added benefits of being provably consistent, robust, and interpretable. Open-source Python implementations of the signal subgraph estimator, along with all of our methods, are freely available online \cite{chungGraSPyGraphStatistics2019a}.

Collectively, the proposed methods presented in this work motivate a number of potential future extensions. First, there is room to develop statistical innovations that will be able to determine if a particular random graph model appropriately models a given real-world dataset. This can be accomplished via tests of goodness-of-fit (GoF), an important criterion for model selection \cite{MYUNG2000190}. A GoF test has already been developed for the inhomogeneous Erd\H{o}s--R\'enyi random graph \cite{pmlr-v119-dan20a}, a particular version of the \ie model, and similar tests can be developed for the \rdpg (vertex-level model) and the \sbm (community-level model). Once developed, these tests can be used as a validation step before running any of our proposed multi-level tests, adding further rigor to these methods.

Second, while \Dcorr enables statistically powerful edge-wise testing, it is very computationally expensive. Performing $k$-sample testing with \Dcorr typically requires a costly permutation test to estimate the null distribution and subsequent p-value. While there is a good chi-square approximation for the null distribution of unbiased \Dcorr with comparable finite-sample power \cite{shenChiSquareTestDistance2020}, this test generally requires a sample size $\geq 20$ to be statistically valid. In connectomics, it can sometimes be difficult to achieve a sample this large, particularly if one is studying a rare neurological disorder or using a very time-intensive process to estimate the connectome. 

Third, these methods are designed for the comparison of samples of connectomes from distinct categorical or dimensional phenotypes. Statistical modeling of the connectome in relation to a continuous phenotypic variable of interest (such as age) is a fundamental challenge for the analysis of dynamic connectomes. While our methods can be applied if the continuous variable is discretized into categorical bins, extensions are required to enable true regression analysis of the connectome in this statistical modeling framework. 

All statistical methods described in this paper are implemented in \texttt{graspologic}, an open-source Python package for statistical network analysis maintained by Microsoft Research (\url{https://github.com/microsoft/graspologic}). The code necessary to reproduce the analyses, simulations, and figures presented in this work are available in a series of Jupyter Notebooks at (\url{https://github.com/neurodata/MCC}).

\section{Conclusion}
The network-level view of brain organization provided by the connectome will enable a transformative understanding of the brain \cite{morganWhyNotConnectomics2013}. For an organ system whose function---both in disease and in health---remains so poorly understood, the promise of this new data type is immense. However, our ability to \textit{map} connectomes is quickly outpacing our ability to \textit{analyze} them.  As the wealth of neuroimaging and connectomics data continues to grow, new mathematical and statistical techniques will be required to discover the brain circuits that underlie neurological processes, disorders, and diseases. We anticipate that the multi-level algorithms and techniques presented in this work will be widely used by future researchers to uncover the neurobiological correlates of different phenotypes in multi-subject connectomics datasets.

\appendix
\section{Simulations}
\subsection{Edge simulations.}
\label{sec:edgesim}
We consider two populations of networks generated from a two-community \sbm.
Edge weights are sampled from a truncated normal distribution to emulate correlation matrices.
All networks have $n=20$ vertices with $10$ vertices belonging to the first community and $10$ vertices belonging to the second community. The community probability matrix for each population is given by
\begin{align*}
    \B^{1} &= 
    \begin{bmatrix}
        \tnorm(0, 0.25) & \tnorm(0, 0.25) \\
        \tnorm(0, 0.25) & \tnorm(0, 0.25) 
    \end{bmatrix} \\
    \B^{2} &= 
    \begin{bmatrix}
        \tnorm(0 + \delta, 0.25 + \phi) & \tnorm(0, 0.25) \\
        \tnorm(0, 0.25) & \tnorm(0, 0.25) 
    \end{bmatrix}
\end{align*}
where $\tnorm(\mu, \sigma^2)$ denotes a truncated normal distribution with mean $\mu$ and variance $\sigma^2$ such that all values are bounded within $[-1, 1]$. 
A total of $m = 20, 40, \dots, 1000$ networks are sampled ($m/2$ networks per population). 
In the first population, all edges are sampled from the same edge weight distribution $\tnorm(0, 0.25)$.
In the second population, all edge weights are also sampled from $\tnorm(0, 0.25)$ except for those in the first community.
In this community, the distribution of edge weights has either a different mean, $\delta$, or variance, $0.25 + \phi$, from the first population.
Therefore, the edges in the first community of these simulated connectomes are the signal edges that we hope to correctly identify. 
For each edge, p-values are computed by three different tests: 1) \ttest, 2) Mann-Whitney (\mw) U test, a non-parametric test of medians, and 3) 2-sample Distance Correlation (\Dcorr) test, a test of equality in distribution. 
For each test, the p-values are sorted to find the ten most significant edges, and the performance is evaluated with precision (the number of signal edges that are correctly identified).

Figure \ref{fig:edgesim} shows the Precision@10 as a function of sample size, mean, and variance. 
Figure \ref{fig:edgesim} top row shows that all three tests can identify signal edges that are different in mean, and that no particular test is superior than another in this setting. 
Figure \ref{fig:edgesim} bottom row shows that only \Dcorr can detect signal edges with differences in variance when the means are kept the same. 
This is because \ttest and \mw test for differences in location (e.g., mean or median), whereas \Dcorr tests for any differences between a pair of observed distributions.
Because \Dcorr achieves about the same precision and recall as the location tests when only location varies, and demonstrates considerably better operating characteristics when the variance varies, we use it in the real-world data.
Note that when looking at the top-$k$ edges, the choice of $k$ is arbitrary.
The salient point from this simulation is that if we look at the top-10 edges identified by each algorithm, none of those edges selected by the \ttest and \mw are signal edges --- on average, they are false positives.

If the sample size $m \geq 20$, one can use a chi-square test that well approximates the $k$-sample \Dcorr test \cite{shenChiSquareTestDistance2020}. This can improve the computational efficiency of edge-wise testing by avoiding the costly permutation test that \Dcorr normally uses to estimate the null distribution.

\subsection{Vertex simulations.}
\label{sec:nodesim}
One goal of connectomics is to identify signal vertices that are different between populations. 
In this section, we identify signal vertices using different vector-based vertex representations. 
The first representation we consider is the simplest: it is possible describe a vertex using all its incident edges via the corresponding row (or column) of a vertex in the adjacency matrices, and compare these vectors using Multivariate Distance Matrix Regression (MDMR) \cite{kimComparisonStatisticalTests2014}. 
The second representation we consider, which is highly popular in modern connectomics literature, is a set of vertex-level graph statistics. Specifically, we fit an exponential random graph model (ERGM) \cite{simpsonExponentialRandomGraph2011} with the following graph statistics for each vertex: local clustering coefficient (LCC), betweenness centrality (BC), closeness centrality (CC), and number of triangles.
The third representation we consider is the network-based statistic (NBS), which identifies signal vertices as those that comprise subgraphs of strong signal edges \cite{zaleskyNetworkbasedStatisticIdentifying2010}.
In the traditional setup for NBS, signal edges are identified using a \ttest (or \anova in multifactorial designs).
Based on the results of our edge-wise simulation (\S\ref{sec:edgesim}), we also test the performance of NBS when the \ttest is substituted for \Dcorr.
The final representation we consider are the low-dimensional latent-space positions estimated by \omni.
Since all vertex representations are multivariate, hypotheses are tested using \hotellings, a multivariate generalization of the \ttest.

We consider a population of \rdpg s in two different settings where the number of signal vertices is varied. In both settings, $m$ null vertices are sampled from the latent position $X_1 = [0.25, 0.25]$ and $n$ signal vertices are sampled from $X_2 = \mathrm{rot}(70^\circ)X_1 \,,$ where $\mathrm{rot}(\theta)$ represents a 2-dimensional rotation matrix of angle $\theta$. In setting 1, $X_2$ is stretched, and a second \rdpg is constructed with latent positions $X_1$ and $X_3 = 0.4 X_2$ and an equivalent number of vertices per each position. In setting 2, $X_2$ is stretched and rotated (i.e., $X_3 = 0.4 \mathrm{rot}(10^\circ)X_2$). The vertices sampled from $X_2$ or $X_3$ (i.e., from different latent positions) are considered signal vertices, and we vary the number of these vertices from $n = 0, 5, 10, 15, 20, 25$. The number of null vertices is set to $m = 50 - n \,.$ A total of $100$ networks are sampled per population, and the p-values are computed using \hotellings on each of the three vertex representations for each vertex. Vertices with p-values less than $\alpha=0.05$ after Holm--Bonferroni correction are classified as signal vertices in this simulation. The performance of each algorithm is measured using a Receiver Operator Characteristic (ROC) curve, which shows the trade-off between the False Positive Rate and True Positive Rate for each method.

Figure \ref{fig:nodesim} shows that \omni is uniformly a more accurate method for identifying signal vertices. Across all settings, its accuracy (measured by the Area Under the ROC curve) is higher than the other three methods. In particular, the ERGM, which uses the widely popular approach of graph statistics, performs much worse at this task.
Substitute the \ttest in NBS with \Dcorr improves the method's ability to detect signal edges by about 3\%, but the performance of both variants lags far behind \omni.

\subsection{Community simulations.}
\label{sec:commsim}
We consider two populations of networks generated from a three-community block diagonal \sbm. As in the edge simulation (Methods \S \ref{sec:edgesim}), edge weights are sampled from truncated normal distributions to emulate correlation matrices. All communities have $n=10$ vertices each, and the community probability matrix for each population is given by
\begin{align*}
    \B^{1} &= 
    \begin{bmatrix}
        \tnorm(0, 0.25) \\
        & \tnorm(0, 0.25) \\
        & & \tnorm(-0.75, 0.25) 
    \end{bmatrix} \,,
\end{align*}
and
\begin{align*}
    \B^{2} &= 
    \begin{bmatrix}
        \tnorm(0, 0.25) \\
        & \tnorm(0, 0.50) \\
        & & \tnorm(0.75, 0.25) 
    \end{bmatrix} \,.
\end{align*}
A total of $m = 10, 20, \dots, 100$ networks are sampled ($m/2$ networks per population). Note that the edge weight distributions in the first diagonal block are equal, while the distributions have different variances in the second diagonal block and different means in the third diagonal block. Therefore, data from block one allow us to measure the false positive rate (FPR), while data from blocks two and three allow us to measure the true positive rate (TPR). Each simulation run was conducted 50 times to ensure accurate estimation of the FPR and TPR. All p-values were corrected using Holm--Bonferroni with significance determined at $\alpha = 0.05$.

From these simulations, we see that all four tests have roughly the same FPR (about 5\% regardless of sample size) (Figure \ref{fig:commsim} \textit{first panel}). However, they differ in their ability to detect signal communities. When all populations have the same mean edge weight, but different variances, only the Multivariate Weighted method consistently detects signal communities once the sample size is greater than $25$ networks (Figure \ref{fig:commsim} \textit{second panel}). All other methods have a lower TPR in this setting. When the edge weights have different means, the TPR of the Multivariate Weighted method lags behind the Average Edge Weight method in low sample size settings ($N<25$) (Figure \ref{fig:commsim} \textit{third panel}). However, when the sample size is larger than $N>25 \,,$ the Multivariate Weighted and Average Edge Weight methods perform comparably. These simulations demonstrate that the Multivariate Weighted signal community detection method, which includes the most information about connectivity within a block, achieves the highest TPR while maintaining a FPR less than or equal to the significance value of the test.

\section{Supplementary investigations of real data}

\paragraph{Visualizations of connectomes.}
The average adjacency matrix of each strain was computed and visualized as a heatmap (Supplementary Figure \ref{fig:connectomes}). Connectomes in this dataset are undirected and parcellated with a bilateral atlas.

\begin{suppfigure*}[!ht]
\includegraphics[width=\linewidth]{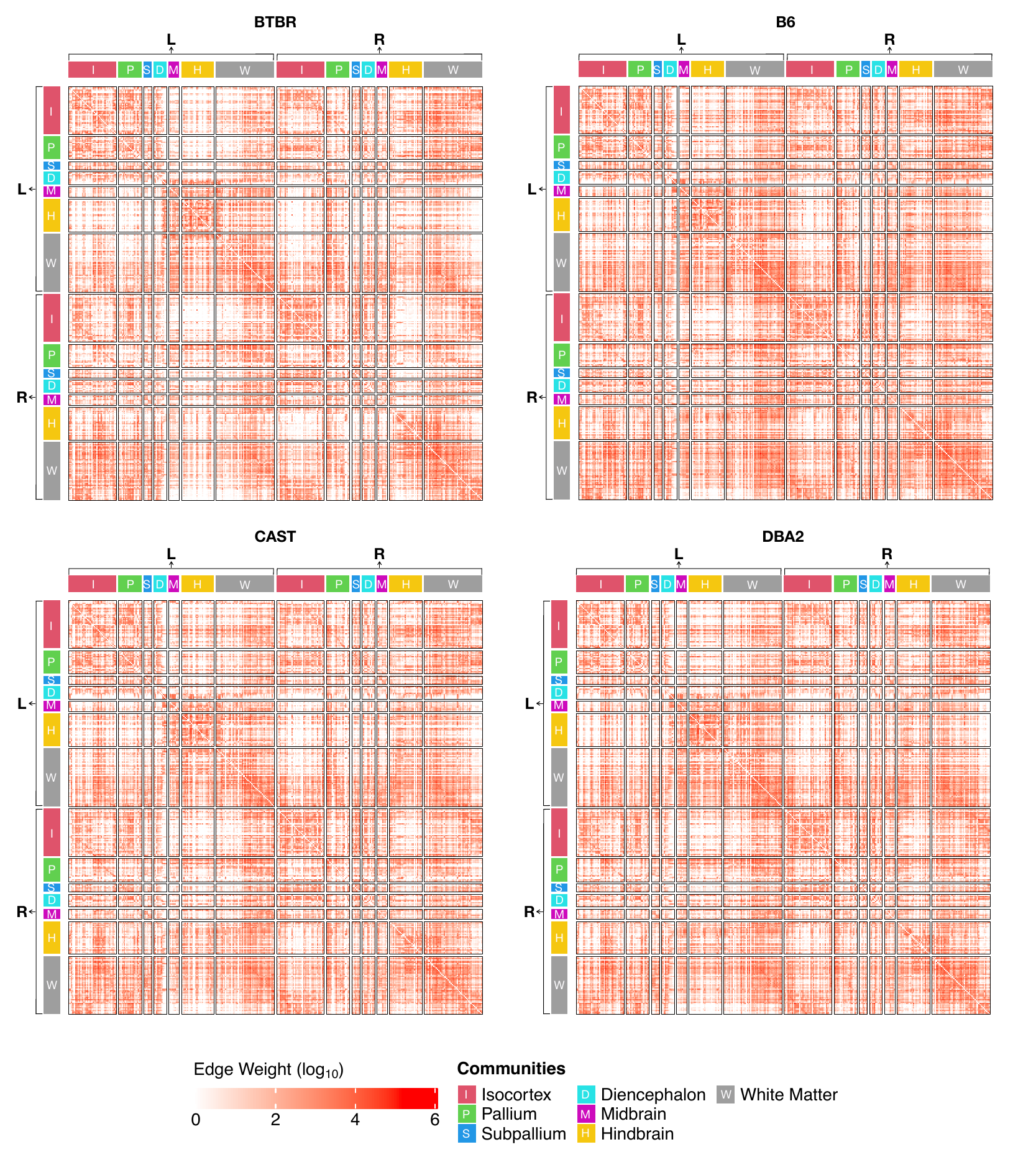}
    \caption{Log-transformed average adjacency matrices for each mouse strain with hierarchical structure labels. Hierarchical labels show the hemispheric and superstructure level. For hemispheric labels, L (R) denotes the left (right) hemisphere. Superstructures are labeled as follows: I) isocortex, P) pallium, S) subpallium, D) diencephalon, M) midbrain, H) hindbrain, and W) white matter.}
\label{fig:connectomes}
\end{suppfigure*} 

\paragraph{Comparison of \textit{mesoscale} methods.}
In Supplementary Figure \ref{fig:mgc}, we show the log-transformed $p$-value for each approach, with communities in blue denoting signal communities (significant after Holm--Bonferroni correction). 

\begin{suppfigure*}[ht!!!]
\centering
\includegraphics[width=0.9\linewidth]{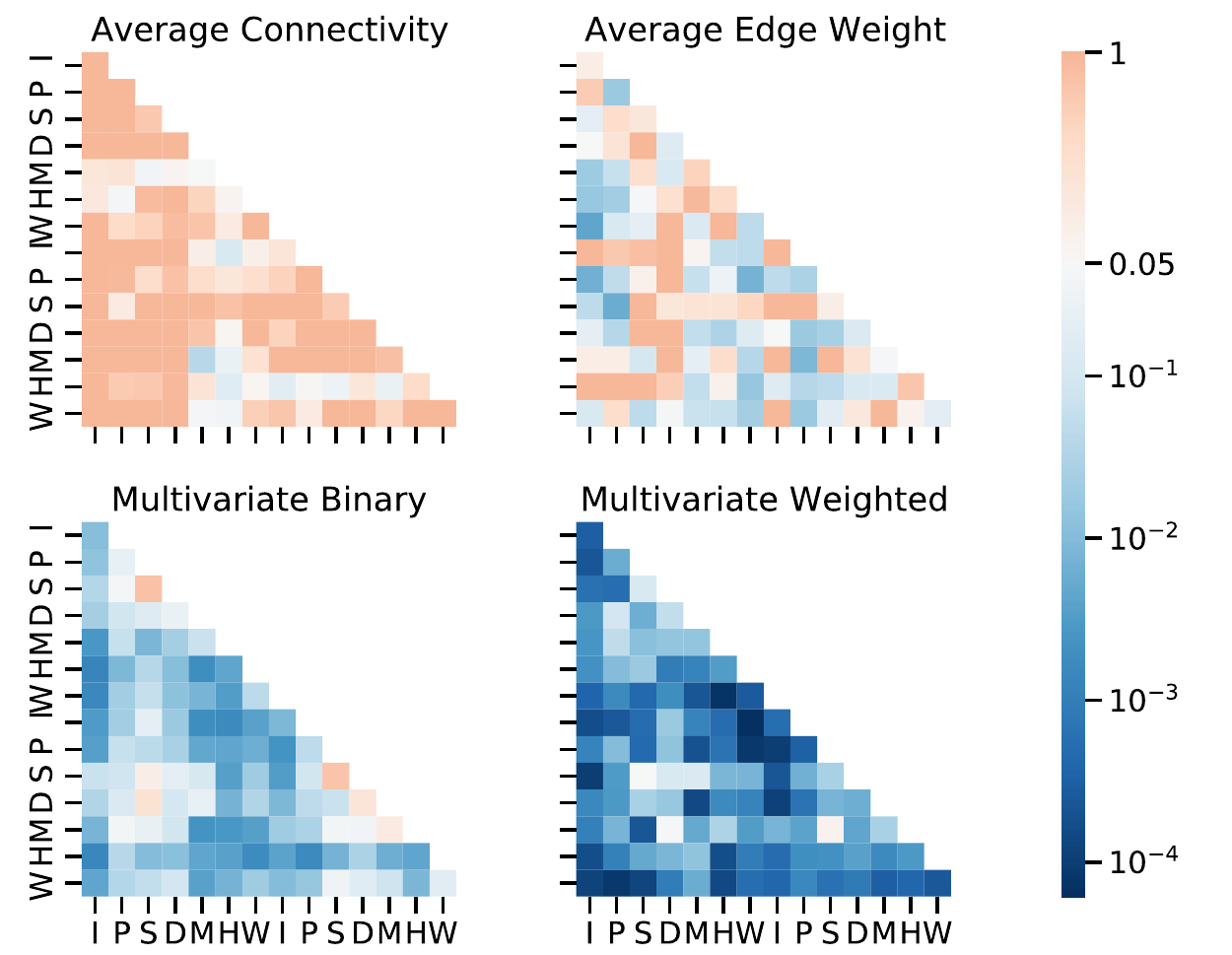}
    \caption{P-values for each proposed approach for summarizing the information in a communities. More signal communities are found as the amount of information encoded by the summary method increases. Because connectomes in this dataset are undirected, this significance matrix will be symmetric and therefore only the upper triangle of these matrices should be considered. The colorbar represents $\log_{10}$ $p$-values. Significance is determined at $\alpha = 0.05$ following Holm--Bonferroni correction; therefore, communities in blue are signal communities.}
\label{fig:mgc}
\end{suppfigure*}

\paragraph{Embeddings of weak signal vertices.}
The strongest signal vertex identified by this method in our real-world data was the left hemisphere corpus callosum. In Figure \ref{fig:corpus}, we demonstrate how a \textit{pairs plot}, a $d$-dimensional scatter plot matrix, can be used to visualize the vertex embedding produced by \omni. In Supplementary Figure \ref{fig:weak_vertex}, we show pairs plots of two weak signal vertices for comparison.

\begin{suppfigure*}[!ht]
\centering
\includegraphics[width=\linewidth]{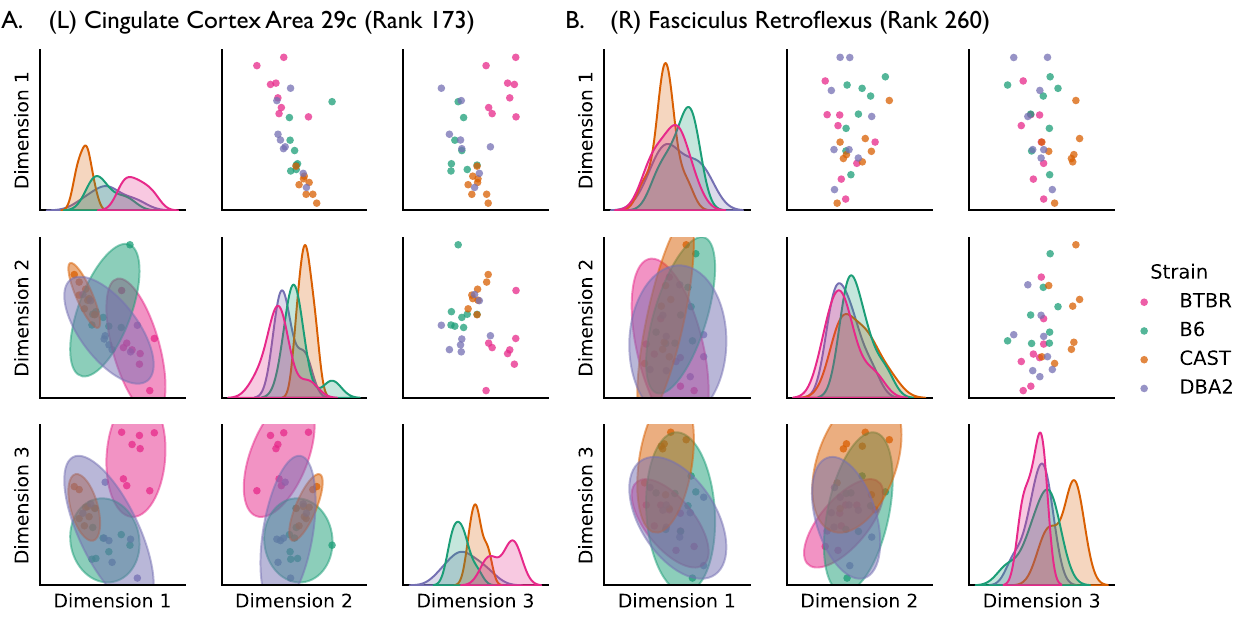}
    \caption{Pairs plots of the vertex embeddings of the left cingulate cortex area 29c and the right fasciculus retroflexus, two weak signal vertices identified by \omni\ and \manova. As shown by the kernel density estimates (\textit{diagonal}) and the 95\% prediction ellipses (\textit{lower triangle}), the distribution of these embeddings is not separable. This emphasizes the homogeneity of these vertices across strains relative to the two strongest signal vertices shown in Figure \ref{fig:corpus}.}
\label{fig:weak_vertex}
\end{suppfigure*}

\paragraph{Visualizing null edges, vertices, and communities.x}
Using the procedures described above, we identified the weakest signal edge (the left frontal cortex to the right temporal association cortex), vertex (the left medial longitudinal fasciculus and tectospinal tract), and community (left isocortex to left diencephalon) across all mouse strains. To contrast with the strongest signal structures shown in Figure \ref{fig:main}, we plotted tractograms of the weakest components neurological structures (Supplementary Figure \ref{fig:sup-main}). These tractograms are much more homogeneous than those plotted in Figure \ref{fig:main}, as are the distributions of numerical features for each component (shown in the bottom row). Since this edge, vertex, and community contain no information about the strain of a mouse, we term them \textit{null components}.

\begin{suppfigure*}[ht!]
\centering
\includegraphics[width=0.74\linewidth]{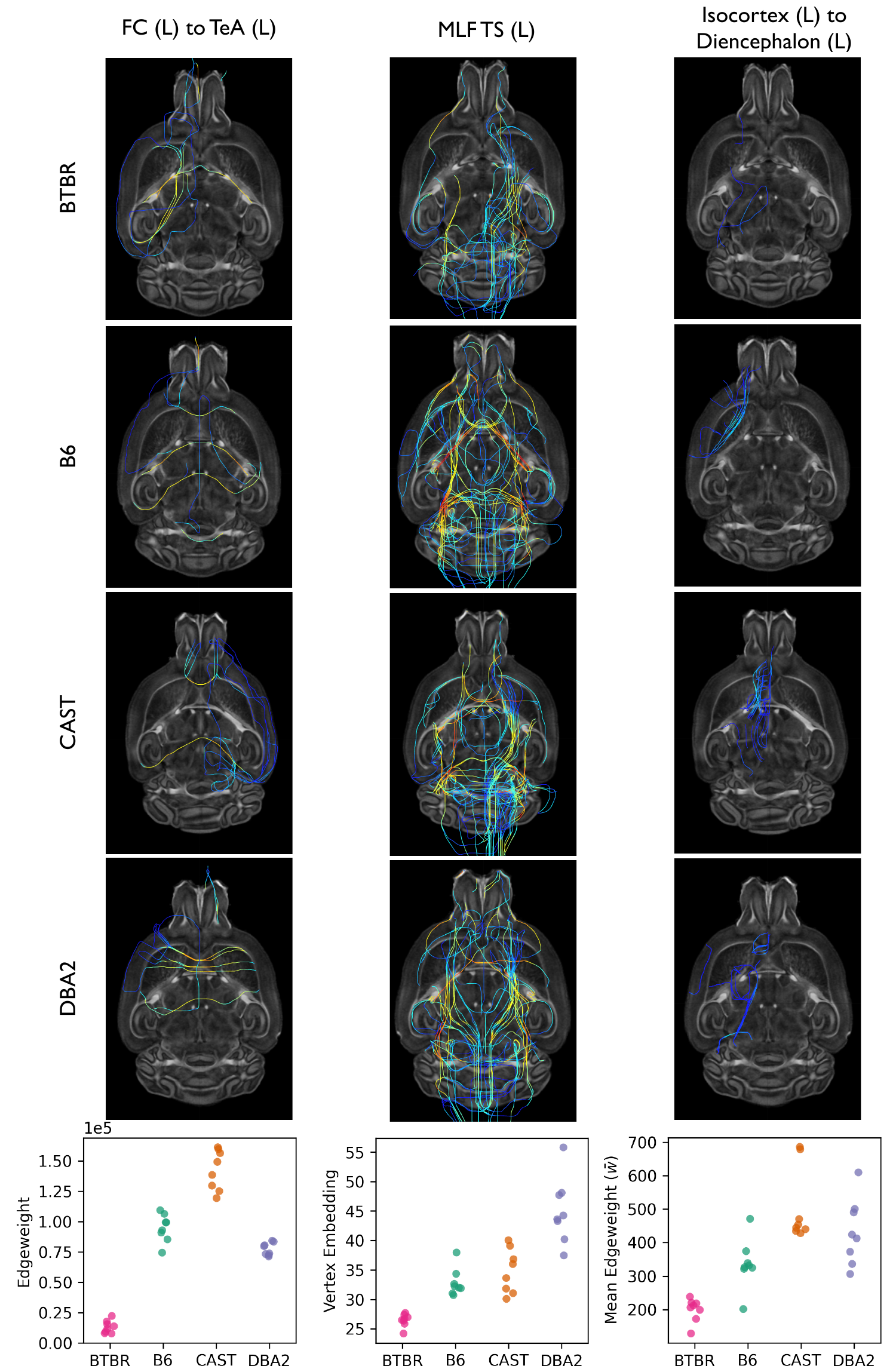}
    \caption{Visualization of the weakest signal edge (the left frontal cortex to the right temporal association cortex), vertex (the left medial longitudinal fasciculus and tectospinal tract), and community (left isocortex to left diencephalon) across all mouse strains. At each topological level, tractograms of these neurological structures are shown for each mouse strain. Compared to Figure \ref{fig:main}, tractograms here are much more homogeneous across strains, displaying minor differences in connectivity. (\textit{Bottom row}) The distribution of edge weights for the weakest signal edge (\textit{Column 1}); the distribution of the first embedding dimension for the weakest signal vertex (\textit{Column 2}); and the distribution of average edge weight for the most weakest signal community (\textit{Column 3}). Each dot represents data from an individual mouse. Differences in distributions are much less pronounced compared to Figure \ref{fig:main}. In these distributions, one strain is typically aberrant while the other three strains are homogeneous.}
\label{fig:sup-main}
\end{suppfigure*}

\begin{table*}[ht!]
\centering
\begin{tabular}{@{}llll@{}}
\toprule
\textbf{Vertex}                & \textbf{statistic} & \textbf{p-value} \\ \midrule
Substantia Nigra (L) & 0.855 & 3.28e-05 \\
Middle Cerebellar Peduncle (R) & 0.852 & 3.45e-05 \\
Internal Capsule (L) & 0.851 & 3.5e-05 \\
Substantia Nigra (R) & 0.845 & 3.84e-05 \\
Pontine Reticular Nucleus (R) & 0.840 & 4.14e-05 \\
Pontine Reticular Nucleus (L) & 0.838 & 4.25e-05 \\
Parasubiculum (L) & 0.838 & 4.25e-05 \\
Ventral Tegmental Area (R) & 0.837 & 4.32e-05 \\
Retro Rubral Field (L) & 0.829 & 4.91e-05 \\
Fastigial Medial Nucleus of Cerebellum (R) & 0.824 & 5.3e-05 \\
Cerebral Peduncle (R) & 0.819 & 5.74e-05 \\
Cerebral Peduncle (L) & 0.818 & 5.87e-05 \\
Fastigial Medial Nucleus of Cerebellum (L) & 0.814 & 6.23e-05 \\
Fimbria (L) & 0.813 & 6.28e-05 \\
Brain Stem Rest (L) & 0.811 & 6.46e-05 \\
Cingulate Cortex Area 30 (L) & 0.810 & 6.6e-05 \\
Ventral Tegmental Area (L) & 0.806 & 6.99e-05 \\
Globus Pallidus (R) & 0.802 & 7.46e-05 \\
Internal Capsule (R) & 0.797 & 8.09e-05 \\
Subthalamic Nucleus (R) & 0.792 & 8.78e-05 \\
Zona Incerta (R) & 0.783 & 0.000102 \\
Parabrachial Nucleus (L) & 0.783 & 0.000102 \\
Globus Pallidus (L) & 0.781 & 0.000104 \\
Insular Cortex (R) & 0.780 & 0.000106 \\
Brain Stem Rest (R) & 0.777 & 0.00011 \\
Secondary Visual Cortex Mediolateral Area (R) & 0.777 & 0.00011 \\
Inferior Colliculus (R) & 0.776 & 0.000111 \\
Superior Cerebellar Peduncle (L) & 0.775 & 0.000113 \\
Retro Rubral Field (R) & 0.773 & 0.000117 \\
Stria Terminalis (L) & 0.772 & 0.000118 \\
Postsubiculum (R) & 0.769 & 0.000123 \\
Midbrain Reticular Nucleus (L) & 0.768 & 0.000125 \\
Striatum (R) & 0.768 & 0.000125 \\
Primary Visual Cortex Binocular Area (R) & 0.768 & 0.000125 \\
Corpus Callosum (L) & 0.767 & 0.000125 \\ \bottomrule
\end{tabular}
\caption{The top 34 signal vertices (out of 326 total vertices) detected by MDMR, ranked by their Holm--Bonferroni corrected p-values. Using MDMR, the corpus callosum is the 34th ranked signal vertex.}
\label{tab:mdmr_vertex}
\end{table*}

\clearpage

\begin{table*}[]
\centering
\begin{tabular}{@{}l|cc|cc|c@{}}
 &
  \multicolumn{2}{l|}{\textit{Left Hemisphere}} &
  \multicolumn{2}{l|}{\textit{Right Hemisphere}} &
  \multicolumn{1}{l}{} \\ \midrule
\textbf{Vertex} &
  \multicolumn{1}{l}{\textbf{p-value}} &
  \multicolumn{1}{l|}{\textbf{Rank}} &
  \multicolumn{1}{l}{\textbf{p-value}} &
  \multicolumn{1}{l|}{\textbf{Rank}} &
  \multicolumn{1}{l}{\textbf{Avg. Rank}} \\ \midrule
Corpus Callosum & 5.09e-25 & 1 & 1.09e-23 & 2 & 1.5 \\
Secondary Motor Cortex & 7.73e-20 & 3 & 3.1e-16 & 7 & 5.0 \\
Internal Capsule & 1.6e-14 & 11 & 3.26e-16 & 8 & 9.5 \\
Stria Terminalis & 1.25e-13 & 18 & 2.59e-14 & 12 & 15.0 \\
Fimbria & 1.94e-19 & 5 & 3.58e-12 & 26 & 15.5 \\
Ventral Tegmental Area & 3.87e-12 & 27 & 8.76e-13 & 21 & 24.0 \\
Hippocampus & 8.02e-11 & 40 & 7.43e-14 & 15 & 27.5 \\
Ectorhinal Cortex & 8.11e-12 & 29 & 4.56e-11 & 37 & 33.0 \\
Globus Pallidus & 9.68e-11 & 41 & 1.89e-12 & 25 & 33.0 \\
Cerebral Peduncle & 4.64e-10 & 57 & 9.98e-15 & 10 & 33.5 \\ \bottomrule
\end{tabular}
\caption{The top 10 bilateral signal vertex pairs (out of 181 total vertex pairs), as determined by \omni and \manova, ranked by their average ranked p-values.}
\label{tab:bilateral-vertices}
\end{table*}

\clearpage
\section{Ethics}
The Duke Institutional Animal Care and Use Committee approved all the study protocols used to acquire the data analyzed in this manuscript.

\section{Author contributions}
\textbf{VG:} Formal analysis; Investigation; Methodology; Software; Validation; Visualization; Writing (original draft, review and editing).
\textbf{JC:} Formal analysis; Software; Visualization; Writing (review and editing).
\textbf{EB:} Formal analysis; Software; Writing (review and editing).
\textbf{BDP:} Formal analysis; Software; Writing (review and editing).
\textbf{JA:} Formal analysis; Writing (review and editing).
\textbf{LU:} Data curation; Project administration; Resources; Writing (review and editing).
\textbf{GAJ:} Data curation; Writing (review and editing).
\textbf{NW:} Data curation; Writing (review and editing).
\textbf{YP:} Software; Data curation; Writing (review and editing).
\textbf{CEP:} Supervision; Investigation; Methodology; Writing (review and editing).
\textbf{JTV:} Supervision; Investigation; Methodology; Conceptualization; Funding acquisition; Writing (review and editing).

\section{Declaration of competing interests}
The authors have no competing interests to declare.

\section{Acknowledgments}
This work was supported by funding from Microsoft Research, and by NIH 1RF1MH121539-01 and R01 AG066184/AG/NIA NIH HHS/United States.

\clearpage
\bibliography{references}

\begin{thebibliography}{63}
\providecommand{\natexlab}[1]{#1}
\providecommand{\url}[1]{\texttt{#1}}
\expandafter\ifx\csname urlstyle\endcsname\relax
  \providecommand{\doi}[1]{doi: #1}\else
  \providecommand{\doi}{doi: \begingroup \urlstyle{rm}\Url}\fi

\bibitem[Anscombe(1973)]{anscombe_graphs_1973}
F.~J. Anscombe.
\newblock Graphs in statistical analysis.
\newblock \emph{The American Statistician}, 27\penalty0 (1):\penalty0 17--21,
  1973.
\newblock ISSN 0003-1305, 1537-2731.
\newblock \doi{10.1080/00031305.1973.10478966}.
\newblock URL
  \url{http://www.tandfonline.com/doi/abs/10.1080/00031305.1973.10478966}.

\bibitem[Athreya et~al.(2017)Athreya, Fishkind, Levin, Lyzinski, Park, Qin,
  Sussman, Tang, Vogelstein, and
  Priebe]{athreyaStatisticalInferenceRandom2017a}
Avanti Athreya, Donniell~E. Fishkind, Keith Levin, Vince Lyzinski, Youngser
  Park, Yichen Qin, Daniel~L. Sussman, Minh Tang, Joshua~T. Vogelstein, and
  Carey~E. Priebe.
\newblock Statistical inference on random dot product graphs: A survey.
\newblock \emph{arXiv:1709.05454 [math, stat]}, September 2017.

\bibitem[Baggio et~al.(2018)Baggio, Abos, Segura, Campabadal, Garcia-Diaz,
  Uribe, Compta, Marti, Valldeoriola, and
  Junque]{baggioStatisticalInferenceBrain2018}
Hugo~C. Baggio, Alexandra Abos, Barbara Segura, Anna Campabadal, Anna
  Garcia-Diaz, Carme Uribe, Yaroslau Compta, Maria~Jose Marti, Francesc
  Valldeoriola, and Carme Junque.
\newblock Statistical inference in brain graphs using threshold-free
  network-based statistics.
\newblock \emph{Human Brain Mapping}, 39\penalty0 (6):\penalty0 2289--2302,
  June 2018.
\newblock ISSN 1065-9471, 1097-0193.
\newblock \doi{10.1002/hbm.24007}.

\bibitem[Benjamini and Hochberg(1995)]{benjaminiControllingFalseDiscovery1995}
Yoav Benjamini and Yosef Hochberg.
\newblock Controlling the {False} {Discovery} {Rate}: {A} {Practical} and
  {Powerful} {Approach} to {Multiple} {Testing}.
\newblock \emph{Journal of the Royal Statistical Society: Series B
  (Methodological)}, 57\penalty0 (1):\penalty0 289--300, 1995.
\newblock ISSN 2517-6161.
\newblock \doi{10.1111/j.2517-6161.1995.tb02031.x}.
\newblock URL
  \url{https://onlinelibrary.wiley.com/doi/abs/10.1111/j.2517-6161.1995.tb02031.x}.
\newblock \_eprint:
  https://onlinelibrary.wiley.com/doi/pdf/10.1111/j.2517-6161.1995.tb02031.x.

\bibitem[Betzel et~al.(2018)Betzel, Medaglia, and
  Bassett]{betzelDiversityMesoscaleArchitecture2018}
Richard~F. Betzel, John~D. Medaglia, and Danielle~S. Bassett.
\newblock Diversity of meso-scale architecture in human and non-human
  connectomes.
\newblock \emph{Nature Communications}, 9\penalty0 (1):\penalty0 346, December
  2018.
\newblock ISSN 2041-1723.
\newblock \doi{10.1038/s41467-017-02681-z}.

\bibitem[Bullmore and Sporns(2009)]{bullmoreComplexBrainNetworks2009}
Ed~Bullmore and Olaf Sporns.
\newblock Complex brain networks: Graph theoretical analysis of structural and
  functional systems.
\newblock \emph{Nature Reviews Neuroscience}, 10\penalty0 (3):\penalty0
  186--198, March 2009.
\newblock ISSN 1471-003X, 1471-0048.
\newblock \doi{10.1038/nrn2575}.

\bibitem[Bullmore and Bassett(2011)]{bullmoreBrainGraphsGraphical2011a}
Edward~T. Bullmore and Danielle~S. Bassett.
\newblock Brain {{Graphs}}: {{Graphical Models}} of the {{Human Brain
  Connectome}}.
\newblock \emph{Annual Review of Clinical Psychology}, 7\penalty0 (1):\penalty0
  113--140, April 2011.
\newblock ISSN 1548-5943, 1548-5951.
\newblock \doi{10.1146/annurev-clinpsy-040510-143934}.

\bibitem[Calabrese et~al.(2015)Calabrese, Badea, Cofer, Qi, and
  Johnson]{calabreseDiffusionMRITractography2015}
Evan Calabrese, Alexandra Badea, Gary Cofer, Yi~Qi, and G.~Allan Johnson.
\newblock A {{Diffusion MRI Tractography Connectome}} of the {{Mouse Brain}}
  and {{Comparison}} with {{Neuronal Tracer Data}}.
\newblock \emph{Cerebral Cortex}, 25\penalty0 (11):\penalty0 4628--4637,
  November 2015.
\newblock ISSN 1047-3211, 1460-2199.
\newblock \doi{10.1093/cercor/bhv121}.

\bibitem[Chen et~al.(2019)Chen, Huroyan, Soni, Lu, Maciejewski, and
  Kobourov]{chenSameStatsDifferent2019}
Hang Chen, Vahan Huroyan, Utkarsh Soni, Yafeng Lu, Ross Maciejewski, and
  Stephen Kobourov.
\newblock Same {{Stats}}, {{Different Graphs}}: {{Exploring}} the {{Space}} of
  {{Graphs}} in {{Terms}} of {{Graph Properties}}.
\newblock \emph{IEEE Transactions on Visualization and Computer Graphics},
  pages 1--1, 2019.
\newblock ISSN 1077-2626, 1941-0506, 2160-9306.
\newblock \doi{10.1109/TVCG.2019.2946558}.

\bibitem[Chung et~al.(2019)Chung, Pedigo, Bridgeford, Varjavand, Helm, and
  Vogelstein]{chungGraSPyGraphStatistics2019a}
Jaewon Chung, Benjamin~D. Pedigo, Eric~W. Bridgeford, Bijan~K. Varjavand,
  Hayden~S. Helm, and Joshua~T. Vogelstein.
\newblock {{GraSPy}}: {{Graph}} statistics in python.
\newblock \emph{Journal of Machine Learning Research}, 20\penalty0
  (158):\penalty0 1--7, 2019.

\bibitem[Chung et~al.(2021)Chung, Bridgeford, Arroyo, Pedigo, Saad-Eldin,
  Gopalakrishnan, Xiang, Priebe, and
  Vogelstein]{chungStatisticalConnectomics2020a}
Jaewon Chung, Eric Bridgeford, Jesús Arroyo, Benjamin~D. Pedigo, Ali
  Saad-Eldin, Vivek Gopalakrishnan, Liang Xiang, Carey~E. Priebe, and Joshua~T.
  Vogelstein.
\newblock Statistical connectomics.
\newblock \emph{Annual Review of Statistics and Its Application}, 8\penalty0
  (1):\penalty0 463--492, 2021.
\newblock \doi{10.1146/annurev-statistics-042720-023234}.
\newblock URL \url{https://doi.org/10.1146/annurev-statistics-042720-023234}.

\bibitem[Cox and Cox(2008)]{coxMultidimensionalScaling2008}
Michael A.~A. Cox and Trevor~F. Cox.
\newblock \emph{Multidimensional {{Scaling}}}, pages 315--347.
\newblock {Springer Berlin Heidelberg}, {Berlin, Heidelberg}, 2008.
\newblock ISBN 978-3-540-33036-3 978-3-540-33037-0.
\newblock \doi{10.1007/978-3-540-33037-0_14}.

\bibitem[Craddock et~al.(2013)Craddock, Jbabdi, Yan, Vogelstein, Castellanos,
  Di~Martino, Kelly, Heberlein, Colcombe, and
  Milham]{craddockImagingHumanConnectomes2013a}
R~Cameron Craddock, Saad Jbabdi, Chao-Gan Yan, Joshua~T Vogelstein, F~Xavier
  Castellanos, Adriana Di~Martino, Clare Kelly, Keith Heberlein, Stan Colcombe,
  and Michael~P Milham.
\newblock Imaging human connectomes at the macroscale.
\newblock \emph{Nature Methods}, 10\penalty0 (6):\penalty0 524--539, June 2013.
\newblock ISSN 1548-7091, 1548-7105.
\newblock \doi{10.1038/nmeth.2482}.

\bibitem[Craddock et~al.(2015)Craddock, Tungaraza, and
  Milham]{craddockConnectomicsNewApproaches2015a}
R~Cameron Craddock, Rosalia~L Tungaraza, and Michael~P Milham.
\newblock Connectomics and new approaches for analyzing human brain functional
  connectivity.
\newblock \emph{GigaScience}, 4\penalty0 (1):\penalty0 13, December 2015.
\newblock ISSN 2047-217X.
\newblock \doi{10.1186/s13742-015-0045-x}.

\bibitem[Dagenbach(2019)]{dagenbachInsightsCognitionNetwork2019}
Dale Dagenbach.
\newblock Insights into cognition from network science analyses of human brain
  functional connectivity: {{Working}} memory as a test case.
\newblock In \emph{Connectomics}, pages 27--41. {Elsevier}, 2019.
\newblock ISBN 978-0-12-813838-0.
\newblock \doi{10.1016/B978-0-12-813838-0.00002-9}.

\bibitem[Dan and Bhattacharya(2020)]{pmlr-v119-dan20a}
Soham Dan and Bhaswar~B. Bhattacharya.
\newblock Goodness-of-fit tests for inhomogeneous random graphs.
\newblock In Hal~Daumé III and Aarti Singh, editors, \emph{Proceedings of the
  37th International Conference on Machine Learning}, volume 119 of
  \emph{Proceedings of Machine Learning Research}, pages 2335--2344. PMLR,
  13--18 Jul 2020.
\newblock URL \url{http://proceedings.mlr.press/v119/dan20a.html}.

\bibitem[Efron(2008)]{efronSimultaneousInferenceWhen2008}
Bradley Efron.
\newblock Simultaneous inference: {{When}} should hypothesis testing problems
  be combined?
\newblock \emph{The Annals of Applied Statistics}, 2\penalty0 (1):\penalty0
  197--223, March 2008.
\newblock ISSN 1932-6157.
\newblock \doi{10.1214/07-AOAS141}.

\bibitem[Ellegood et~al.(2013)Ellegood, Babineau, Henkelman, Lerch, and
  Crawley]{ellegoodNeuroanatomicalAnalysisBTBR2013a}
Jacob Ellegood, Brooke~A. Babineau, R.~Mark Henkelman, Jason~P. Lerch, and
  Jacqueline~N. Crawley.
\newblock Neuroanatomical analysis of the {{BTBR}} mouse model of autism using
  magnetic resonance imaging and diffusion tensor imaging.
\newblock \emph{NeuroImage}, 70:\penalty0 288--300, April 2013.
\newblock ISSN 10538119.
\newblock \doi{10.1016/j.neuroimage.2012.12.029}.

\bibitem[Emerson et~al.(2013)Emerson, Green, Schloerke, Crowley, Cook, Hofmann,
  and Wickham]{emersonGeneralizedPairsPlot2013}
John~W. Emerson, Walton~A. Green, Barret Schloerke, Jason Crowley, Dianne Cook,
  Heike Hofmann, and Hadley Wickham.
\newblock The {{Generalized Pairs Plot}}.
\newblock \emph{Journal of Computational and Graphical Statistics}, 22\penalty0
  (1):\penalty0 79--91, January 2013.
\newblock ISSN 1061-8600, 1537-2715.
\newblock \doi{10.1080/10618600.2012.694762}.

\bibitem[{Fang-Cheng Yeh} et~al.(2010){Fang-Cheng Yeh}, Wedeen, and
  Tseng]{fang-chengyehGeneralizedSamplingImaging2010}
{Fang-Cheng Yeh}, Van~Jay Wedeen, and Wen-Yih~Isaac Tseng.
\newblock Generalized \$\{ Q\}\$-{{Sampling Imaging}}.
\newblock \emph{IEEE Transactions on Medical Imaging}, 29\penalty0
  (9):\penalty0 1626--1635, September 2010.
\newblock ISSN 0278-0062, 1558-254X.
\newblock \doi{10.1109/TMI.2010.2045126}.

\bibitem[Fornito et~al.(2013)Fornito, Zalesky, and
  Breakspear]{fornitoGraphAnalysisHuman2013}
Alex Fornito, Andrew Zalesky, and Michael Breakspear.
\newblock Graph analysis of the human connectome: {{Promise}}, progress, and
  pitfalls.
\newblock \emph{NeuroImage}, 80:\penalty0 426--444, October 2013.
\newblock ISSN 10538119.
\newblock \doi{10.1016/j.neuroimage.2013.04.087}.

\bibitem[Gu et~al.(2016)Gu, Eils, and Schlesner]{guComplexHeatmapsReveal2016a}
Zuguang Gu, Roland Eils, and Matthias Schlesner.
\newblock Complex heatmaps reveal patterns and correlations in multidimensional
  genomic data.
\newblock \emph{Bioinformatics}, 32\penalty0 (18):\penalty0 2847--2849,
  September 2016.
\newblock ISSN 1367-4803, 1460-2059.
\newblock \doi{10.1093/bioinformatics/btw313}.

\bibitem[Hoff et~al.(2002)Hoff, Raftery, and
  Handcock]{hoffLatentSpaceApproaches2002}
Peter~D Hoff, Adrian~E Raftery, and Mark~S Handcock.
\newblock Latent {{Space Approaches}} to {{Social Network Analysis}}.
\newblock \emph{Journal of the American Statistical Association}, 97\penalty0
  (460):\penalty0 1090--1098, December 2002.
\newblock ISSN 0162-1459, 1537-274X.
\newblock \doi{10.1198/016214502388618906}.

\bibitem[Holland et~al.(1983)Holland, Laskey, and
  Leinhardt]{hollandStochasticBlockmodelsFirst1983a}
Paul~W. Holland, Kathryn~Blackmond Laskey, and Samuel Leinhardt.
\newblock Stochastic blockmodels: {{First}} steps.
\newblock \emph{Social Networks}, 5\penalty0 (2):\penalty0 109--137, June 1983.
\newblock ISSN 03788733.
\newblock \doi{10.1016/0378-8733(83)90021-7}.

\bibitem[Holm(1979)]{holmSimpleSequentiallyRejective1979}
Sture Holm.
\newblock A {{Simple Sequentially Rejective Multiple Test Procedure}}.
\newblock \emph{Scandinavian Journal of Statistics}, 6\penalty0 (2):\penalty0
  65--70, 1979.
\newblock ISSN 0303-6898.

\bibitem[Johnson et~al.(2010)Johnson, Badea, Brandenburg, Cofer, Fubara, Liu,
  and Nissanov]{johnsonWaxholmSpaceImagebased2010}
G.~Allan Johnson, Alexandra Badea, Jeffrey Brandenburg, Gary Cofer, Boma
  Fubara, Song Liu, and Jonathan Nissanov.
\newblock Waxholm {{Space}}: {{An}} image-based reference for coordinating
  mouse brain research.
\newblock \emph{NeuroImage}, 53\penalty0 (2):\penalty0 365--372, November 2010.
\newblock ISSN 10538119.
\newblock \doi{10.1016/j.neuroimage.2010.06.067}.

\bibitem[Kaiser(2011)]{kaiserTutorialConnectomeAnalysis2011}
Marcus Kaiser.
\newblock A tutorial in connectome analysis: {{Topological}} and spatial
  features of brain networks.
\newblock \emph{NeuroImage}, 57\penalty0 (3):\penalty0 892--907, August 2011.
\newblock ISSN 10538119.
\newblock \doi{10.1016/j.neuroimage.2011.05.025}.

\bibitem[Kiar et~al.(2017)Kiar, Bridgeford, Gray~Roncal, {Consortium for
  Reliability and Reproducibility (CoRR)}, Chandrashekhar, Mhembere, Ryman,
  Zuo, Margulies, Craddock, Priebe, Jung, Calhoun, Caffo, Burns, Milham, and
  Vogelstein]{kiarHighThroughputPipelineIdentifies2017}
Gregory Kiar, Eric~W. Bridgeford, William~R. Gray~Roncal, {Consortium for
  Reliability and Reproducibility (CoRR)}, Vikram Chandrashekhar, Disa
  Mhembere, Sephira Ryman, Xi-Nian Zuo, Daniel~S. Margulies, R.~Cameron
  Craddock, Carey~E. Priebe, Rex Jung, Vince~D. Calhoun, Brian Caffo, Randal
  Burns, Michael~P. Milham, and Joshua~T. Vogelstein.
\newblock A {{High}}-{{Throughput Pipeline Identifies Robust Connectomes But
  Troublesome Variability}}.
\newblock Preprint, {Neuroscience}, September 2017.

\bibitem[Kim et~al.(2014)Kim, Wozniak, Mueller, Shen, and
  Pan]{kimComparisonStatisticalTests2014}
Junghi Kim, Jeffrey~R. Wozniak, Bryon~A. Mueller, Xiaotong Shen, and Wei Pan.
\newblock Comparison of statistical tests for group differences in brain
  functional networks.
\newblock \emph{NeuroImage}, 101:\penalty0 681--694, November 2014.
\newblock ISSN 10538119.
\newblock \doi{10.1016/j.neuroimage.2014.07.031}.

\bibitem[Levin et~al.(2017)Levin, Athreya, Tang, Lyzinski, and
  Priebe]{levinCentralLimitTheorem2017b}
Keith Levin, Avanti Athreya, Minh Tang, Vince Lyzinski, and Carey~E. Priebe.
\newblock A {{Central Limit Theorem}} for an {{Omnibus Embedding}} of
  {{Multiple Random Dot Product Graphs}}.
\newblock In \emph{2017 {{IEEE International Conference}} on {{Data Mining
  Workshops}} ({{ICDMW}})}, pages 964--967, {New Orleans, LA}, November 2017.
  {IEEE}.
\newblock ISBN 978-1-5386-3800-2.
\newblock \doi{10.1109/ICDMW.2017.132}.

\bibitem[McFarlane et~al.(2008)McFarlane, Kusek, Yang, Phoenix, Bolivar, and
  Crawley]{mcfarlaneAutismlikeBehavioralPhenotypes2008}
H.~G. McFarlane, G.~K. Kusek, M.~Yang, J.~L. Phoenix, V.~J. Bolivar, and J.~N.
  Crawley.
\newblock Autism-like behavioral phenotypes in {{BTBR T}}+tf/{{J}} mice.
\newblock \emph{Genes, Brain and Behavior}, 7\penalty0 (2):\penalty0 152--163,
  March 2008.
\newblock ISSN 1601-1848, 1601-183X.
\newblock \doi{10.1111/j.1601-183X.2007.00330.x}.

\bibitem[Meyza and Blanchard(2017)]{meyzaBTBRMouseModel2017}
K.Z. Meyza and D.C. Blanchard.
\newblock The {{BTBR}} mouse model of idiopathic autism \textendash{}
  {{Current}} view on mechanisms.
\newblock \emph{Neuroscience \& Biobehavioral Reviews}, 76:\penalty0 99--110,
  May 2017.
\newblock ISSN 01497634.
\newblock \doi{10.1016/j.neubiorev.2016.12.037}.

\bibitem[Morgan and Lichtman(2013)]{morganWhyNotConnectomics2013}
Joshua~L Morgan and Jeff~W Lichtman.
\newblock Why not connectomics?
\newblock \emph{Nature Methods}, 10\penalty0 (6):\penalty0 494--500, June 2013.
\newblock ISSN 1548-7091, 1548-7105.
\newblock \doi{10.1038/nmeth.2480}.

\bibitem[Myers et~al.(2019)Myers, Arvapalli, Ramachandran, Pisner, Frank,
  Lemmer, Bridgeford, Nikolaidis, and
  Vogelstein]{myersStandardizingHumanBrain2019}
Patrick~E. Myers, Ganesh~C. Arvapalli, Sandhya~C. Ramachandran, Derek~A.
  Pisner, Paige~F. Frank, Allison~D. Lemmer, Eric~W. Bridgeford, Aki
  Nikolaidis, and Joshua~T. Vogelstein.
\newblock Standardizing {{Human Brain Parcellations}}.
\newblock Preprint, {Neuroscience}, November 2019.

\bibitem[Myung(2000)]{MYUNG2000190}
In~Jae Myung.
\newblock The importance of complexity in model selection.
\newblock \emph{Journal of Mathematical Psychology}, 44\penalty0 (1):\penalty0
  190--204, 2000.
\newblock ISSN 0022-2496.
\newblock \doi{https://doi.org/10.1006/jmps.1999.1283}.
\newblock URL
  \url{https://www.sciencedirect.com/science/article/pii/S002224969991283X}.

\bibitem[Nenning et~al.(2020)Nenning, Xu, Schwartz, Arroyo, Woehrer, Franco,
  Vogelstein, Margulies, Liu, Smallwood, Milham, and
  Langs]{nenningJointEmbeddingScalable2020}
Karl-Heinz Nenning, Ting Xu, Ernst Schwartz, Jesus Arroyo, Adelheid Woehrer,
  Alexandre~R. Franco, Joshua~T. Vogelstein, Daniel~S. Margulies, Hesheng Liu,
  Jonathan Smallwood, Michael~P. Milham, and Georg Langs.
\newblock Joint embedding: {{A}} scalable alignment to compare individuals in a
  connectivity space.
\newblock \emph{NeuroImage}, 222:\penalty0 117232, November 2020.
\newblock ISSN 10538119.
\newblock \doi{10.1016/j.neuroimage.2020.117232}.

\bibitem[Otsu(1979)]{otsuThresholdSelectionMethod1979}
Nobuyuki Otsu.
\newblock A {{Threshold Selection Method}} from {{Gray}}-{{Level Histograms}}.
\newblock \emph{IEEE Transactions on Systems, Man, and Cybernetics}, 9\penalty0
  (1):\penalty0 62--66, January 1979.
\newblock ISSN 0018-9472, 2168-2909.
\newblock \doi{10.1109/TSMC.1979.4310076}.

\bibitem[Panda et~al.(2020)Panda, Palaniappan, Xiong, Bridgeford, Mehta, Shen,
  and Vogelstein]{pandaHyppoComprehensiveMultivariate2020}
Sambit Panda, Satish Palaniappan, Junhao Xiong, Eric~W. Bridgeford, Ronak
  Mehta, Cencheng Shen, and Joshua~T. Vogelstein.
\newblock Hyppo: {{A Comprehensive Multivariate Hypothesis Testing Python
  Package}}.
\newblock \emph{arXiv:1907.02088 [cs, stat]}, August 2020.

\bibitem[Rubinov and Sporns(2010)]{rubinovComplexNetworkMeasures2010}
Mikail Rubinov and Olaf Sporns.
\newblock Complex network measures of brain connectivity: {{Uses}} and
  interpretations.
\newblock \emph{NeuroImage}, 52\penalty0 (3):\penalty0 1059--1069, September
  2010.
\newblock ISSN 10538119.
\newblock \doi{10.1016/j.neuroimage.2009.10.003}.

\bibitem[Scattoni et~al.(2008)Scattoni, Gandhy, Ricceri, and
  Crawley]{scattoniUnusualRepertoireVocalizations2008}
Maria~Luisa Scattoni, Shruti~U. Gandhy, Laura Ricceri, and Jacqueline~N.
  Crawley.
\newblock Unusual {{Repertoire}} of {{Vocalizations}} in the {{BTBR T}}+tf/{{J
  Mouse Model}} of {{Autism}}.
\newblock \emph{PLoS ONE}, 3\penalty0 (8):\penalty0 e3067, August 2008.
\newblock ISSN 1932-6203.
\newblock \doi{10.1371/journal.pone.0003067}.

\bibitem[Scheinerman and Tucker(2010)]{scheinermanModelingGraphsUsing2010}
Edward~R. Scheinerman and Kimberly Tucker.
\newblock Modeling graphs using dot product representations.
\newblock \emph{Computational Statistics}, 25\penalty0 (1):\penalty0 1--16,
  March 2010.
\newblock ISSN 0943-4062, 1613-9658.
\newblock \doi{10.1007/s00180-009-0158-8}.

\bibitem[Shen and
  Vogelstein(2020{\natexlab{a}})]{shenChiSquareTestDistance2020}
Cencheng Shen and Joshua~T. Vogelstein.
\newblock The {{Chi}}-{{Square Test}} of {{Distance Correlation}}.
\newblock \emph{arXiv:1912.12150 [cs, math, stat]}, February
  2020{\natexlab{a}}.

\bibitem[Shen and
  Vogelstein(2020{\natexlab{b}})]{shenExactEquivalenceDistance2020}
Cencheng Shen and Joshua~T. Vogelstein.
\newblock The {{Exact Equivalence}} of {{Distance}} and {{Kernel Methods}} for
  {{Hypothesis Testing}}.
\newblock \emph{arXiv:1806.05514 [cs, stat]}, September 2020{\natexlab{b}}.

\bibitem[Silverman et~al.(2010)Silverman, Tolu, Barkan, and
  Crawley]{silvermanRepetitiveSelfGroomingBehavior2010}
Jill~L Silverman, Seda~S Tolu, Charlotte~L Barkan, and Jacqueline~N Crawley.
\newblock Repetitive {{Self}}-{{Grooming Behavior}} in the {{BTBR Mouse Model}}
  of {{Autism}} is {{Blocked}} by the {{mGluR5 Antagonist MPEP}}.
\newblock \emph{Neuropsychopharmacology}, 35\penalty0 (4):\penalty0 976--989,
  March 2010.
\newblock ISSN 0893-133X, 1740-634X.
\newblock \doi{10.1038/npp.2009.201}.

\bibitem[Simpson et~al.(2011)Simpson, Hayasaka, and
  Laurienti]{simpsonExponentialRandomGraph2011}
Sean~L. Simpson, Satoru Hayasaka, and Paul~J. Laurienti.
\newblock Exponential {{Random Graph Modeling}} for {{Complex Brain Networks}}.
\newblock \emph{PLoS ONE}, 6\penalty0 (5):\penalty0 e20039, May 2011.
\newblock ISSN 1932-6203.
\newblock \doi{10.1371/journal.pone.0020039}.

\bibitem[Sporns et~al.(2005)Sporns, Tononi, and
  K{\"o}tter]{spornsHumanConnectomeStructural2005}
Olaf Sporns, Giulio Tononi, and Rolf K{\"o}tter.
\newblock The {{Human Connectome}}: {{A Structural Description}} of the {{Human
  Brain}}.
\newblock \emph{PLoS Computational Biology}, 1\penalty0 (4):\penalty0 e42,
  2005.
\newblock ISSN 1553-734X, 1553-7358.
\newblock \doi{10.1371/journal.pcbi.0010042}.

\bibitem[Sz{\'e}kely et~al.(2007)Sz{\'e}kely, Rizzo, and
  Bakirov]{szekelyMeasuringTestingDependence2007}
G{\'a}bor~J. Sz{\'e}kely, Maria~L. Rizzo, and Nail~K. Bakirov.
\newblock Measuring and testing dependence by correlation of distances.
\newblock \emph{The Annals of Statistics}, 35\penalty0 (6):\penalty0
  2769--2794, December 2007.
\newblock ISSN 0090-5364.
\newblock \doi{10.1214/009053607000000505}.

\bibitem[{van den Heuvel} et~al.(2016){van den Heuvel}, Bullmore, and
  Sporns]{vandenheuvelComparativeConnectomics2016a}
Martijn~P. {van den Heuvel}, Edward~T. Bullmore, and Olaf Sporns.
\newblock Comparative {{Connectomics}}.
\newblock \emph{Trends in Cognitive Sciences}, 20\penalty0 (5):\penalty0
  345--361, May 2016.
\newblock ISSN 13646613.
\newblock \doi{10.1016/j.tics.2016.03.001}.

\bibitem[Varoquaux and
  Craddock(2013)]{varoquauxLearningComparingFunctional2013}
Ga{\"e}l Varoquaux and R.~Cameron Craddock.
\newblock Learning and comparing functional connectomes across subjects.
\newblock \emph{NeuroImage}, 80:\penalty0 405--415, October 2013.
\newblock ISSN 10538119.
\newblock \doi{10.1016/j.neuroimage.2013.04.007}.

\bibitem[Vogelstein et~al.(2013{\natexlab{a}})Vogelstein, Roncal, Vogelstein,
  and Priebe]{vogelsteinGraphClassificationUsing2013}
J.~T. Vogelstein, W.~G. Roncal, R.~J. Vogelstein, and C.~E. Priebe.
\newblock Graph {{Classification Using Signal}}-{{Subgraphs}}: {{Applications}}
  in {{Statistical Connectomics}}.
\newblock \emph{IEEE Transactions on Pattern Analysis and Machine
  Intelligence}, 35\penalty0 (7):\penalty0 1539--1551, July 2013{\natexlab{a}}.
\newblock ISSN 0162-8828, 2160-9292.
\newblock \doi{10.1109/TPAMI.2012.235}.

\bibitem[Vogelstein et~al.(2013{\natexlab{b}})Vogelstein, Gray~Roncal,
  Vogelstein, and Priebe]{6341752}
Joshua~T. Vogelstein, William Gray~Roncal, R.~Jacob Vogelstein, and Carey~E.
  Priebe.
\newblock Graph classification using signal-subgraphs: Applications in
  statistical connectomics.
\newblock \emph{IEEE Transactions on Pattern Analysis and Machine
  Intelligence}, 35\penalty0 (7):\penalty0 1539--1551, 2013{\natexlab{b}}.
\newblock \doi{10.1109/TPAMI.2012.235}.

\bibitem[Vogelstein et~al.(2019)Vogelstein, Bridgeford, Pedigo, Chung, Levin,
  Mensh, and Priebe]{vogelsteinConnectalCodingDiscovering2019a}
Joshua~T Vogelstein, Eric~W Bridgeford, Benjamin~D Pedigo, Jaewon Chung, Keith
  Levin, Brett Mensh, and Carey~E Priebe.
\newblock Connectal coding: Discovering the structures linking cognitive
  phenotypes to individual histories.
\newblock \emph{Current Opinion in Neurobiology}, 55:\penalty0 199--212, April
  2019.
\newblock ISSN 09594388.
\newblock \doi{10.1016/j.conb.2019.04.005}.

\bibitem[Wang et~al.(2015)Wang, Wang, Xia, Liao, Evans, and
  He]{wangGRETNAGraphTheoretical2015}
Jinhui Wang, Xindi Wang, Mingrui Xia, Xuhong Liao, Alan Evans, and Yong He.
\newblock {{GRETNA}}: A graph theoretical network analysis toolbox for imaging
  connectomics.
\newblock \emph{Frontiers in Human Neuroscience}, 9, June 2015.
\newblock ISSN 1662-5161.
\newblock \doi{10.3389/fnhum.2015.00386}.

\bibitem[Wang et~al.(2020)Wang, Anderson, Ashbrook, Gopalakrishnan, Park,
  Priebe, Qi, Laoprasert, Vogelstein, Williams, and
  Johnson]{wangVariabilityHeritabilityMouse2020a}
Nian Wang, Robert~J Anderson, David~G Ashbrook, Vivek Gopalakrishnan, Youngser
  Park, Carey~E Priebe, Yi~Qi, Rick Laoprasert, Joshua~T Vogelstein, Robert~W
  Williams, and G~Allan Johnson.
\newblock Variability and {{Heritability}} of {{Mouse Brain Structure}}:
  {{Microscopic MRI Atlases}} and {{Connectomes}} for {{Diverse Strains}}.
\newblock \emph{NeuroImage}, page 117274, August 2020.
\newblock ISSN 10538119.
\newblock \doi{10.1016/j.neuroimage.2020.117274}.

\bibitem[Wang et~al.(2018)Wang, Shen, Badea, Priebe, and
  Vogelstein]{wangSignalSubgraphEstimation2018}
Shangsi Wang, Cencheng Shen, Alexandra Badea, Carey~E. Priebe, and Joshua~T.
  Vogelstein.
\newblock Signal {{Subgraph Estimation Via Vertex Screening}}.
\newblock \emph{arXiv:1801.07683 [stat]}, January 2018.

\bibitem[Wang et~al.(2017)Wang, Wen, Pan, and
  Huang]{wangSureIndependenceScreening2017a}
Xueqin Wang, Canhong Wen, Wenliang Pan, and Mian Huang.
\newblock Sure {{Independence Screening Adjusted}} for {{Confounding
  Covariates}} with {{Ultrahigh}}-dimensional {{Data}}.
\newblock \emph{Statistica Sinica}, 2017.
\newblock ISSN 10170405.
\newblock \doi{10.5705/ss.202014.0117}.

\bibitem[Weisstein(2004)]{weissteinBonferroniCorrectionMathWorld2004}
Eric~W. Weisstein.
\newblock Bonferroni {{Correction}}. \{\vphantom\}{{From MathWorld}}---{{A
  Wolfram Web Resource}}\vphantom\{\}, 2004.

\bibitem[Xu et~al.(2020)Xu, Nenning, Schwartz, Hong, Vogelstein, Goulas, Fair,
  Schroeder, Margulies, Smallwood, Milham, and
  Langs]{xuCrossspeciesFunctionalAlignment2020}
Ting Xu, Karl-Heinz Nenning, Ernst Schwartz, Seok-Jun Hong, Joshua~T.
  Vogelstein, Alexandros Goulas, Damien~A. Fair, Charles~E. Schroeder,
  Daniel~S. Margulies, Jonny Smallwood, Michael~P. Milham, and Georg Langs.
\newblock Cross-species functional alignment reveals evolutionary hierarchy
  within the connectome.
\newblock \emph{NeuroImage}, 223:\penalty0 117346, December 2020.
\newblock ISSN 10538119.
\newblock \doi{10.1016/j.neuroimage.2020.117346}.

\bibitem[Yeh and Tseng(2011)]{yehNTU90HighAngular2011a}
Fang-Cheng Yeh and Wen-Yih~Isaac Tseng.
\newblock {{NTU}}-90: {{A}} high angular resolution brain atlas constructed by
  q-space diffeomorphic reconstruction.
\newblock \emph{NeuroImage}, 58\penalty0 (1):\penalty0 91--99, September 2011.
\newblock ISSN 10538119.
\newblock \doi{10.1016/j.neuroimage.2011.06.021}.

\bibitem[Yeh et~al.(2013)Yeh, Verstynen, Wang, Fernández-Miranda, and
  Tseng]{yeh_deterministic_2013}
Fang-Cheng Yeh, Timothy~D. Verstynen, Yibao Wang, Juan~C. Fernández-Miranda,
  and Wen-Yih~Isaac Tseng.
\newblock Deterministic {Diffusion} {Fiber} {Tracking} {Improved} by
  {Quantitative} {Anisotropy}.
\newblock \emph{PLoS ONE}, 8\penalty0 (11):\penalty0 e80713, November 2013.
\newblock ISSN 1932-6203.
\newblock \doi{10.1371/journal.pone.0080713}.
\newblock URL \url{https://dx.plos.org/10.1371/journal.pone.0080713}.

\bibitem[Yeh et~al.(2017)Yeh, Liu, Hitchens, and Wu]{yehMappingImmuneCell2017}
Fang-Cheng Yeh, Li~Liu, T.~Kevin Hitchens, and Yijen~L. Wu.
\newblock Mapping immune cell infiltration using restricted diffusion {{MRI}}:
  {{Restricted Diffusion Imaging}}.
\newblock \emph{Magnetic Resonance in Medicine}, 77\penalty0 (2):\penalty0
  603--612, February 2017.
\newblock ISSN 07403194.
\newblock \doi{10.1002/mrm.26143}.

\bibitem[Yoo et~al.(2017)Yoo, Lee, Chung, Sohn, Chung, Na, Ju, and
  Jeong]{yooDegreeBasedStatistic2017}
Kwangsun Yoo, Peter Lee, Moo~K. Chung, William~S. Sohn, Sun~Ju Chung, Duk~L.
  Na, Daheen Ju, and Yong Jeong.
\newblock Degree-based statistic and center persistency for brain connectivity
  analysis.
\newblock \emph{Human Brain Mapping}, 38\penalty0 (1):\penalty0 165--181,
  January 2017.
\newblock ISSN 1065-9471, 1097-0193.
\newblock \doi{10.1002/hbm.23352}.

\bibitem[Zalesky et~al.(2010)Zalesky, Fornito, and
  Bullmore]{zaleskyNetworkbasedStatisticIdentifying2010}
Andrew Zalesky, Alex Fornito, and Edward~T. Bullmore.
\newblock Network-based statistic: {{Identifying}} differences in brain
  networks.
\newblock \emph{NeuroImage}, 53\penalty0 (4):\penalty0 1197--1207, December
  2010.
\newblock ISSN 10538119.
\newblock \doi{10.1016/j.neuroimage.2010.06.041}.

\end{thebibliography}
\end{document}